\documentclass[aps,prl,twocolumn,groupedaddress,showpacs,amsmath,amssymb]{revtex4-1}
\usepackage{graphicx,color}
\usepackage{epstopdf}
\usepackage{amsmath}
\usepackage{bm}

\begin{document}
\def\ldotsplus{\mathinner{\ldotp\ldotp\ldotp\ldotp}}
\def\fourdots{\relax\ifmmode\ldotsplus\else$\m@th \ldotsplus\,$\fi}

\title{Enhancement of microorganism swimming speed  in active matter}

\author{Harsh Soni$^1$}
\author{Robert A. Pelcovits$^2$}
\author{Thomas R. Powers$^{1,2}$}

\affiliation{$^1$School of Engineering, Brown University, Providence, RI 02912 USA}
\affiliation{$^2$Department of Physics, Brown University, Providence, RI 02012 USA}

\date{\today}

 % no more than 600 characters
\begin{abstract}
We study a swimming undulating sheet in the isotropic phase of an active nematic liquid crystal. Activity changes the effective shear viscosity, reducing it to zero at a critical value of activity. Expanding in the sheet amplitude, we find that the correction to the swimming speed due to activity is inversely proportional to the effective shear viscosity. Our perturbative calculation becomes invalid near the critical value of activity; using numerical methods to probe this regime, we find that activity enhances the swimming speed by an order of magnitude compared to the passive case.
\end{abstract}


\maketitle
\begin{figure}[t]
\includegraphics[width=%0.45\textwidth
	3.3in]{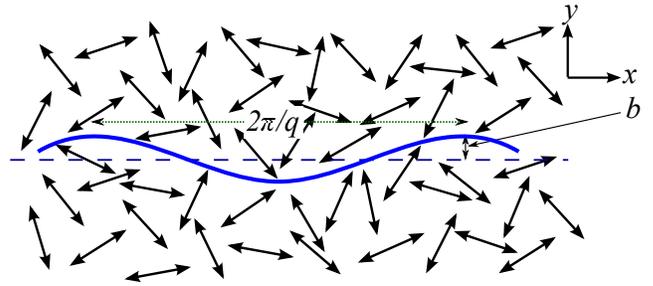}
	%\centering
\caption{(Color online.) %Schematic diagram of a 
A Taylor sheet (blue wavy line) with wavenumber $q$ and amplitude $b$ swimming in an active nematic liquid fluid in the isotropic phase. 
%The blue wavy line is the Taylor sheet and 
The double headed arrows are the active nematic molecules. %The dashed line represents the sheet without a wave. The swimming sheet has wavenumber $q$ and amplitude $b$.
}
\label{setupfig}
\end{figure}

%\section{Introduction}
Recent years have seen many advances in the study of swimming at the micron scale in viscous fluids~\cite{LaugaPowers2009}, such as the creation of artificial microswimmers~\cite{Ishiyama_etal2001,dreyfus05_nature,GhoshFischer2009}, measurements of the flows induced by single 
%and multiple swimmers~\cite{mendelson99,wu00,dombrowski04,sokolov07}, 
swimmers~\cite{Guasto:2010dz,Drescheretal2011,Goldstein2015}, and the development of  hydrodynamic theories~\cite{stone_samuel1996,najafi05,PakLauga2015} and simulations~\cite{pozrikidis1992,Cortez2001,cortez2}. The field has expanded to include swimmers in non-Newtonian fluids, such as viscoelastic polymer solutions~\cite{lauga2007,FuPowersWolgemuth2007,TeranFauciShelley2010,ShenArratia2011,LiuPowersBreuer2011,SpagnolieBook2015}
%, shear-thinning fluids~\cite{RodrigoLauga2013,Montenegro-Jetal2013,GagnonKeimArratia2014,datt2015}, 
%fluids with interfaces~\cite{trouilloud08,WangArdekani2013,MirbagheriFu2016,ReighLauga2017}, 
and liquid crystals~\cite{Zhou_etal2013,MushenheimEtAl2013,KriegerSpagnoliePowers2014}. All of these studies involve passive fluids, in which the energy that drives the flow is added by the internal motors of the swimmer or an external source such as a rotating magnetic field.  In active fluids, on the other hand, the energy that drives the flow is added to the system at the level of the microscopic constituents of the fluid~\cite{srimrmp}. For example, a suspension of  molecular motors and cytoskeletal filaments shows spontaneous flows due to the consumption of ATP in the suspension by the molecular motors~\cite{Schaller2010,Sanchez2012}. It is natural to ask if an active fluid can do work on a swimmer, causing it to swim faster than it would in a passive fluid with the same stroke. In this Letter we investigate this question with the Taylor model of a waving sheet~\cite{taylor1951} in the isotropic phase of an active nematic liquid crystal (Fig~\ref{setupfig}). 

We use the Taylor sheet because it is one of the simplest models for a flagellated swimmer for which analytical calculations of swimming speed are possible. The isotropic state of the fluid is also chosen for simplicity. Below a critical activity, the undisturbed stable state of the active liquid crystal is isotropic with no flow. The motion of a swimmer induces flows around the swimmer which in turn lead to local order; the simple nature of the base state allows us to treat the swimmer problem perturbatively.  An unconfined active nematic in the nematic phase is unstable to spontaneous flow at any value of activity~\cite{Giomi_etal2012}, making an analytic approach  difficult. 

\begin{figure}[t]
	\includegraphics[width=%0.45\textwidth
	3.3in]{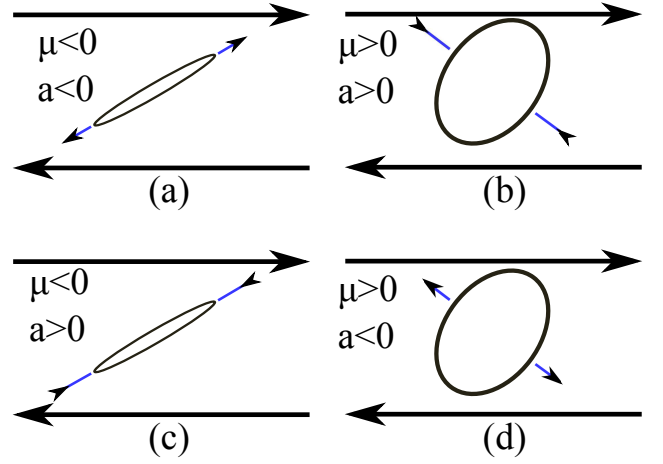}
	%\centering
	\caption{(Color online.)  %Schematic diagram representing the 
	Effect of activity $a$ on the effective shear viscosity of the nematic fluid. The large arrows represent the shear flow and the small arrows represent the %direction of the 
	forces %density 
	due to activity.
	Extensile prolate particles (a) and contractile oblate particles (b) reduce the shear viscosity~\cite{Hatwalne2004}. (Note that the force axis always assumed along the particle axis of symmetry.)  Contractile prolate particles (c) and extensile oblate particles (d) increase the shear viscosity~\cite{Hatwalne2004}.	}
	\label{viscosity}
\end{figure}

We model the isotropic phase of an active nematic by adding activity to de Gennes' hydrodynamic model~\cite{DeGennes1969,degennesbook,deGennes1971} for the isotropic phase of a passive nematic fluid. The governing equations are similar to those used in other studies of active matter~\cite{Hatwalne2004,SaintillanShelley2008,WoodhouseGoldstein2012}. A striking feature of the active isotropic phase of extensile prolate particles (or contractile oblate particles) is that activity reduces the effective shear viscosity (Fig.~\ref{viscosity})~\cite{Hatwalne2004}. In fact, numerical and experimental studies have given evidence for a ``superfluid" state in which the apparent viscosity vanishes for sufficiently large activity~\cite{Cates_etal2008,Sokolov2009,GiomiLiverpoolMarchetti2010,Gachelin2013,Lopez2015}. We find that the swimming speed for a small-amplitude Taylor sheet in our active medium is inversely proportional to the effective shear viscosity. Since our perturbative calculation breaks down when the effective shear viscosity gets too small, we use numerical finite-element methods to show that the swimming speed for small effective viscosity can be an order of magnitude larger than the speed in a passive medium for the same stroke. The outline for the remainder of the Letter is as follows. After introducing the governing equations, we find the critical value of the activity at which the quiescent isotropic state becomes unstable. Then we calculate the swimming speed using perturbation theory, which is valid for a value of activity that is {sufficiently smaller than} the critical value for instability. Finally we numerically calculate the flow, order parameter field, and swimming speed, again assuming the activity is such that the quiescent isotropic state is stable.

To motivate the governing equations, we begin with the nematic degrees of freedom.
For simplicity we assume a one-dimensional deformation of the sheet, with no variation in the spatial direction perpendicular to the plane of the Fig.~\ref{setupfig}. Thus the  the local nematic ordering is characterized  by a symmetric traceless order parameter tensor $Q_{\alpha\beta}$, with  $\alpha, \beta = x,y$. To leading order in $Q_{\alpha\beta}$,  the Landau-de Gennes free energy density is~\cite{degennesbook}
\begin{equation}
\mathcal{F}=\frac{A}{2} Q_{\alpha\beta}Q_{\alpha\beta},\label{free_energy}
\end{equation}
where we sum over repeated indices and $A>0$ in the isotropic phase. Frank elasticity can be neglected in the isotropic phase as can higher order terms in  $Q_{\alpha\beta}$  (note that a cubic term is identically zero in two dimensions). Strictly speaking, a quartic term should be included since the perturbative calculation of the swimming speed requires an expansion to second order in the swimmer amplitude. But the qualitative effect of retaining this term in the calculation is only to slightly change the shape of the potential defined by $\mathcal{F}$, leading to a slight change in the numerical factors in the expression for swimming speed.
Thus the molecular field is $\Phi_{\alpha\beta}\equiv-\partial \mathcal{F}/\partial Q_{\alpha\beta}=-AQ_{\alpha\beta}$ in the isotropic phase. The equilibrium stress is the Ericksen stress, $\sigma^\mathrm{e}_{\alpha\beta}=\mathcal{F}\delta_{\alpha\beta}-\partial{\mathcal F}/\partial(\partial_\beta Q_{\mu\nu})\partial_\alpha Q_{\mu\nu}$~\cite{DeGennes1969,JulicherGrillSalbreaux2018}.

The rate of entropy production per volume is~\cite{degennesbook}
\begin{equation}
T\dot{S}=\sigma'_{\alpha\beta}e_{\alpha\beta}+\Phi_{\alpha\beta}R_{\alpha\beta},
\end{equation}
where $T$ is temperature, $S$ is entropy per volume, $\sigma'_{\alpha\beta}$  is the viscous stress tensor, $e_{\alpha\beta}=(\partial_\alpha v_\beta+\partial_\beta v_\alpha)/2$ is the strain rate tensor, $v_\alpha$ is the velocity field, and  $R_{\alpha\beta}$  is the rate of change of $Q_{\alpha\beta}$ relative to the local rate of rotation $\omega_{\alpha\beta}=(\partial_\alpha v_\beta-\partial_\beta v_\alpha)/2$ of the background fluid,
\begin{equation}
R_{\alpha\beta}=\partial_t Q_{\alpha\beta}+\mathbf{v}\cdot\boldsymbol{\nabla}Q_{\alpha\beta}+\omega_{\alpha\gamma}Q_{\gamma\beta}-Q_{\alpha\gamma}\omega_{\gamma\beta}.
\end{equation}
Following de Gennes~\cite{DeGennes1969}, we take the forces in the entropy source to be the molecular field $\Phi_{\alpha\beta}$  and the viscous stress tensor $\sigma'_{\alpha\beta}$, 
and the corresponding fluxes to be   $e_{\alpha\beta}$ and $R_{\alpha\beta}$. 
Assuming that the forces are linear functions of the fluxes, the phenomenological equations relating the forces to the fluxes are
\begin{eqnarray}
\sigma'_{\alpha\beta}&=&2\eta e_{\alpha\beta}+2(\mu+\mu_1) R_{\alpha\beta}+aQ_{\alpha\beta}\label{stressa}\\
\Phi_{\alpha\beta}&=&2\mu e_{\alpha\beta}+\nu R_{\alpha\beta},\label{Qeq}
\end{eqnarray}
%with the inequality, $\eta  \nu -2 \mu ^2>0$, imposed by the necessity of positive $\dot{S}$~\cite{onsager}. 
where $\eta$ is the shear viscosity,  $\mu$ and $\mu_1$ couple shear and alignment, and $\nu$ is the rotational viscosity. Note that $\eta$, $\mu$, $\mu_1$, and $\nu$ have units of viscosity, and $a$ and $A$ have units of a modulus. We neglect higher-order terms such as $Q_{\alpha\gamma}e_{\gamma\delta}Q_{\delta\beta}$ since the magnitude of the order parameter is small in the isotropic phase. The coefficients $\mu_1$ and $a$ arise from activity. 
When $a=0$ and $\mu_1=0$,  the Onsager reciprocal relations~\cite{onsager} hold, and the rate of entropy production is positive, implying $\eta\nu-2\mu^2>0$. Thus, the active parameter $\mu_1$ determines the degree of violation of the Onsager relations, and, when it is sufficiently positive, can lead to a negative rate of entropy production.  

The active stress is $aQ_{\alpha\beta}$~\cite{Hatwalne2004},  with  $a<0$ for extensile particles and $a>0$ for contractile particles. The coupling $\mu$ controls the orientation of the particles in shear flow, leading to shear birefringence. For example, nematic order develops in a  weak steady shear flow, with $Q_{\alpha\beta}=-(2\mu/A)e_{\alpha\beta}$ to first order in the strain rate~\cite{DeGennes1969}. Note that 
independent of the value of $\mu_1$,  
 particles with  $\mu<0$, such as prolate ellipsoidal particles, align along the shear flow, and particles with  $\mu>0$, such as oblate ellipsoidal particles, align opposite to the shear flow (Fig.~\ref{viscosity}). 

The governing equations are the director equation Eq.~(\ref{Qeq}) and the force balance equation
$\partial_\beta\sigma_{\alpha\beta}=0$,
with  $\sigma_\mathrm{\alpha\beta}=-p\delta_{\alpha\beta}+\sigma^\mathrm{e}_{\alpha\beta}+\sigma'_{\alpha\beta}$. 
We define the effective viscosity $\eta_\mathrm{eff}$ and the effective coupling $\mu_\mathrm{eff}$ by using Eq.~\eqref{Qeq} to eliminate $Q_{\alpha\beta}$ from  the stress, Eq.~\eqref{stressa}, to find %, the viscous stress tensor is given by:
$\sigma'_{\alpha\beta}=2\eta_\text{eff} e_{\alpha\beta}+2 \mu_\text{eff} R_{\alpha\beta}$,
where
\begin{eqnarray}
\eta_\text{eff}&= &\eta-\dfrac{\mu a}{A}\label{etaeff}\\
\mu_\text{eff}&=& \mu+\mu_1-\dfrac{\nu a}{2A}.
\end{eqnarray}
Thus, activity gives rise to an effective shear viscosity $\eta_\text{eff}$ which vanishes at a critical value of the activity $a_\mathrm{c}=A\eta/\mu$.

Next we turn to the linear stability analysis of the state with $v_\alpha=0$ and $Q_{\alpha\beta}=0$, with no swimmer or other confining boundaries. To linear order, the force balance equation  is
\begin{eqnarray}\label{eqfb}
-\partial_\alpha p+ 2\eta_\text{eff}\partial_\beta e_{\alpha\beta}+2 \mu_\text{eff} \partial_\beta {\dot Q}_{\alpha\beta}&=&0,
\end{eqnarray}
where ${\dot Q}_{\alpha\beta}=\partial_tQ_{\alpha\beta}$. 
The pressure $p$ is determined by the  incompressibility constraint,  $\partial_\alpha v_\alpha=0$. It is convenient to enforce incompressibilty with the stream function $\psi$, defined so that $\mathbf{v}=\boldsymbol{\nabla}\times\psi\hat{\mathbf z}$.  Also, in two dimensions, the tensor order parameter $Q_{\alpha\beta}$ is related to the scalar order parameter $S$ and the director $\mathbf{n}$ via $Q_{\alpha\beta}=S(2n_{\alpha}n_\beta-\delta_{\alpha\beta})$.  The linearized equations for the stream function and the order parameter are
\begin{eqnarray}
-\Delta^2\psi+2\frac{\mu_\mathrm{eff}}{\eta_\mathrm{eff}}\left[(\partial_x^2-\partial_y^2){\dot Q}_{xy}-2\partial_x\partial_y{\dot Q}_{xx}\right]&=&0\label{psilin}\\
2\mu\partial_x\partial_y\psi+AQ_{xx}+\nu{\dot Q}_{xx}&=&0\label{Qxxlin}\\
-\mu(\partial_x^2-\partial_y^2)\psi+AQ_{xy}+\nu{\dot Q}_{xy}&=&0,\label{Qxylin}
\end{eqnarray}
where $\nabla=\partial_x^2+\partial_y^2$ . 

For perturbations of the velocity and order parameter tensor proportional to $\exp[\mathrm{i}(\mathbf{q}\cdot\mathbf{x}-\sigma t)]$,
the characteristic equation for this problem yields two roots~\cite{supp},
\begin{eqnarray}
\sigma_1&=&-A/\nu\\
\sigma_2&=&-A \eta_\mathrm{eff}/(\eta_\mathrm{eff}\nu-2\mu\mu_\mathrm{eff}).\label{sigma2}
\end{eqnarray}
The roots are independent of the direction of $\mathbf{q}$ since the base state is isotropic. There are only two roots since the assumption of zero Reynolds number has eliminated $\partial_t v_\alpha$ from the governing equations. The first root $\sigma_1$ is always positive; inserting $\exp[\mathrm{i}(\mathbf{q}\cdot\mathbf{x}-\sigma_1 t)]$ into Eq.~\eqref{Qxxlin} or Eq.~\eqref{Qxylin} reveals that this  mode has no flow, with the director $\mathbf{n}$ parallel to $\mathbf{q}$ for all $\mathbf{x}$ and all $t$,  and the scalar order parameter relaxing to zero with rate $A/\nu$. 

\begin{figure}[t]
\includegraphics[width=3.1in]{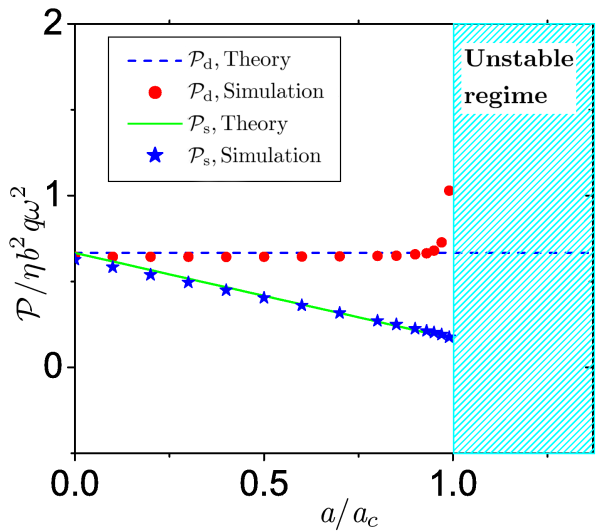}
\caption{(Color online.) Dimensionless rate of working $\cal{P}_\text{s}$ of the swimmer (green line from theory, Eq.~\ref{PS}, blue stars from simulations) vs. dimensionless activity $a/a_\mathrm{c}$, and dimensionless rate of dissipation of energy  $\cal{P}_\text{d}$ (blue dashed line from theory, red dots from simulation) vs. dimensionless activity $a/a_\mathrm{c}$. The parameters used are $\epsilon=bq=0.1$, $\mu=\eta=\nu/3$, $A=\nu \omega$ and $\mu_1=0$. }
\label{power}
\end{figure}

The second root corresponds to a mode in which there is a shear flow with the velocity perpendicular to $\mathbf{q}$ (due to incompressibility $\mathbf{q}\cdot\mathbf{v}=0$), with $\mathbf{n}$ along the flow.  The numerator in Eq.~\eqref{sigma2} is precisely the quantity that determines whether or not the entropy production $T\dot{S}$ is positive. If $\mu_1$ is small enough that $\eta_\mathrm{eff}\nu-2\mu\mu_\mathrm{eff}>0$, then the isotropic state is unstable when $\eta_\mathrm{eff}<0$, i.e. $a> a_\mathrm{c}=A\eta/\mu$ for positive $\mu$ (Fig.~\ref{viscosity}b), or $a<a_\mathrm{c}=A\eta/\mu$ for negative $\mu$ (Fig.~\ref{viscosity}a). The quiescent isotropic state is unstable against shear flow and local ordering when the shear-induced orientation of the particles leads to greater shear flow, as in Figs.~\ref{viscosity}a and~\ref{viscosity}b.
 
We now consider a Taylor swimmer with $y=h(x,t)\equiv b\cos(qx-\omega t)$ (Fig. \ref{setupfig}) in the stable phase of an isotropic active nematic. Our approach is the same as Lauga's calculation for a dilute polymer solution~\cite{lauga2007}. To calculate the swimming speed of the  sheet, we work in the rest frame of the swimmer and solve the governing equations \eqref{stressa} and \eqref{Qeq} with no-slip boundary conditions on the velocity at the swimmer, 
$\mathbf{v}(x, y=h)=\partial_t h(x,t)\hat{\mathbf{y}}$.  The unknown velocity at $y\rightarrow\infty$ is the negative of swimming velocity $U$. No boundary conditions are imposed on the order parameter because we have disregarded the Frank energy. We assume that $\epsilon=b q\ll 1$ and expand in powers of $\epsilon$, so that e.g. $\psi=\epsilon\psi^{(1)}+\epsilon^2\psi^{(2)}$. To first order in $\epsilon$ the equations~(\ref{psilin}--\ref{Qxylin}) yield
\begin{eqnarray}
\psi^{(1)}&=&({\omega}/{q^2})(1+qy)e^{-qy}\cos(q x -\omega t)\label{psifirst}\\
Q^{(1)}_{xx}&=&\dfrac{-2qy\omega\mu e^{-qy} }{A^2+\omega^2\nu^2}\left[\omega\nu\cos(q x -\omega t)+A \sin(q x -\omega t) \right]\nonumber\\
Q^{(1)}_{xy}&=&\dfrac{-2qy\omega\mu e^{-qy} }{A^2+\omega^2\nu^2}\left[A\cos(q x -\omega t) -\omega\nu \sin(q x -\omega t)\right].\nonumber
\end{eqnarray} 
The velocity field is the same as the Stokes flow found by Taylor~\cite{taylor1951} for a Newtonian fluid, and the order parameter is independent of the activity. Note that the direction of $\mathbf{n}$ is independent of $y$ to first order in $\epsilon$, since the ratio $Q_{xy}^{(1)}/Q_{xx}^{(1)}$ is independent of $y$. 

The power $P_\mathrm{s}$ supplied by the swimmer is equal to the sum of the rate of change of the free energy and the net power dissipated in the fluid, $P_\mathrm{s}=\mathrm{d}F/\mathrm{d}t+P_\mathrm{f}$, where $F=\int\mathrm{d}^3x\mathcal{F}$ and
 \begin{eqnarray}
P_\mathrm{s}&=&-\int\left[v_\alpha\sigma_{\alpha\beta}+\frac{\partial\mathcal{F}}{\partial(\partial_\beta Q_{\mu\nu})}\frac{\mathrm{d}Q_{\mu\nu}}{{\mathrm d}t}\right]N_{\beta}\mathrm{d}S\\
P_\mathrm{f}&=&\int\left[e_{\alpha\beta}(\sigma_{\alpha\beta}-\sigma^\mathrm{e}_{\alpha\beta})+\Phi_{\alpha\beta}\frac{\mathrm{d}Q_{\alpha\beta}}{\mathrm{d}t}\right]\mathrm{d}^3x.\label{freeEbalance}
\end{eqnarray}
Here $\mathrm{d}Q_{\alpha\beta}/\mathrm{d}t=\partial_tQ_{\alpha\beta}+v_\gamma\partial_\gamma Q_{\alpha\beta}$, $\mathrm{d}S$ is area element of the swimmer, and $\hat{\mathbf N}$ is the outward-pointing normal to the swimmer. Note that the net power dissipated in the fluid may be negative due to activity. 
%The rate of change of the free energy is equal to the power dissipated in or provided by the active fluid plus the rate of work done by the swimmer~\cite{JulicherGrillSalbreaux2018},
%\begin{eqnarray}
%\frac{{\mathrm d}F}{{\mathrm d}t}&=&\int\left[-e_{\alpha\beta}(\sigma_{\alpha\beta}-\sigma^\mathrm{e}_{\alpha\beta})-\Phi_{\alpha\beta}\frac{\mathrm{d}Q_{\alpha\beta}}{\mathrm{d}t}\right]\mathrm{d}^2x\nonumber\\
%&+&\int\left[v_\alpha\sigma_{\alpha\beta}+\frac{\partial\mathcal{F}}{\partial(\partial_\beta Q_{\mu\nu})}\frac{\mathrm{d}Q_{\mu\nu}}{{\mathrm d}t}\right]N_{\beta}\mathrm{d}s\label{freeEbalance}
%\end{eqnarray}
%where $F=\int\mathcal{F}\mathrm{d}^2x$ is the integral of the free energy density over the time-varying volume of fluid,  $\mathrm{d}Q_{\alpha\beta}/\mathrm{d}t=\partial_tQ_{\alpha\beta}+v_\gamma\partial_\gamma Q_{\alpha\beta}$, $s$ is the arclength parameter along the swimmer, and $\hat{\mathbf N}$ is the outward-pointing normal to the swimmer.
The first-order solutions allow us to calculate the leading order rate of working of the swimmer per unit area of the sheet,
\begin{equation}\mathcal{P}_\mathrm{s}
%-\int\mathrm{d}x\sigma_{\alpha\beta}v_\alpha N_\beta
\approx b^2q\omega^2\left[\eta_\mathrm{eff}-\frac{2\nu\mu\mu_\mathrm{eff}\omega^2}{A^2+\nu^2\omega^2}\right]\label{PS}
\end{equation} 
(note that $P_\mathrm{s}=\int\mathrm{d}S\mathcal{P}_\mathrm{s}$).
The power supplied by the swimmer decreases linearly with activity $a$ (Fig.~\ref{power}, green solid line). The fluid does net positive work on the swimmer when $a>a_0=a_\mathrm{c}+[\eta\nu-2\mu(\mu+\mu_1)]\omega^2/(A\mu)$. The value of $a_0$ can be less than $a_\mathrm{c}$ and  in the regime where our perturbative calculation is valid when $\mu_1$ is sufficiently large and positive. We denote the power  dissipated in the fluid per unit area of the sheet by $\mathcal{P}_\mathrm{d}\equiv \mathcal{P}_\mathrm{f}(a=0,\mu_1=0)$; $\mathcal{P}_\mathrm{d}$ is positive and independent of activity (Fig.~\ref{power}, blue dashed line), and to leading order is given by~\eqref{PS} with  $\eta_\mathrm{eff}$ replaced by $\eta$ and $\mu_\mathrm{eff}$ replaced by $\mu$.

To find the swimming speed, it is convenient to write the time-average of the $x$-component of momentum balance in terms of the velocity and expand to second order in $\epsilon$:
\begin{equation}
\eta_\text{eff}\frac{\mathrm{d}^2}{\mathrm{d}y^2}\langle v_x^{(2)}\rangle+4 e^{-2 qy} (qy-2) yq^2 \omega^3 \dfrac{2 \mu \nu \mu_\text{eff}}{A^2+\nu^2\omega^2}=0.\label{eq2} 
\end{equation}
Enforcing the no-slip boundary condition to second order leads to $\langle v_x^{(2)}(x,0)\rangle=\omega/(2q)$. Solving for the flow leads to the swimming speed

\begin{equation}
U=\frac{c \epsilon^2}{2}\left[1-\frac{2\nu\mu \mu_\text{eff}\omega^2}{\eta_\text{eff}(A^2+\nu^2\omega^2)}\right],\label{Udis0}
\end{equation}
where $c=\omega/q$ is the wave speed of the deformation of the swimmer, and we are using the convention that a positive $U$ means the swimmer moves left in the laboratory frame. In the supplementary material~\cite{supp} we show that  the swimming speed of a two-dimensional squirmer has the same dependence on frequency $\omega$ and material parameters $\nu$, $\mu$, $\mu_\mathrm{eff}$, $\eta_\mathrm{eff}$, and $A$.
The swimming speed diverges when $a\rightarrow a_\mathrm{c}$ since the effective shear viscosity vanishes at the critical activity, indicating a breakdown of the perturbative calculation. Analyzing the form of the next order terms reveals that they are of the order of $\epsilon^4/(a_\mathrm{c}-a)^3$, indicating that the perturbative approach requires $\epsilon^2\ll(a_\mathrm{c}-a)^2$. Also, when $a <a_\mathrm{c}$, $U$ is positive. Thus as long as the fluid is stable, activity cannot make the swimmer swim in the direction of the propagating waves.

\begin{figure}[t]
\includegraphics[width=3.1in]{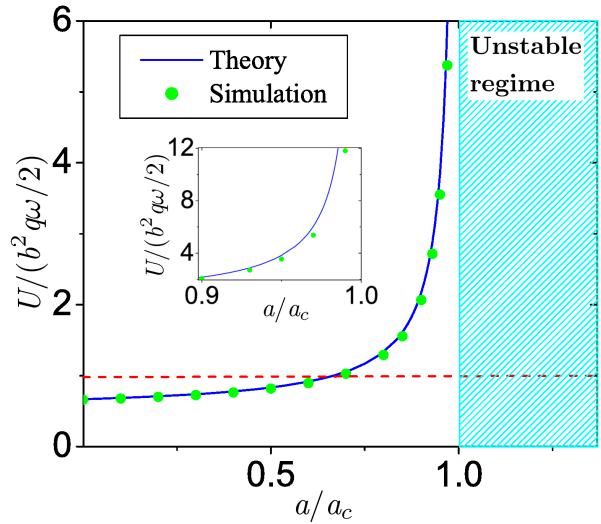}
\caption{(Color online.)  Dimensionless swimming speed $U/(b^2 q\omega/2)$ vs. dimensionless activity $a/a_\mathrm{c}$   from theory, Eq.~\ref{Udis0}, (blue line) and simulations (green dots). The parameters used are the same as in Fig.~\ref{power}. The inset shows the critical region $a\approx a_c$.}
\label{speed}
\end{figure}

\begin{figure}[t]
 	\includegraphics[width=
 	3.1in]{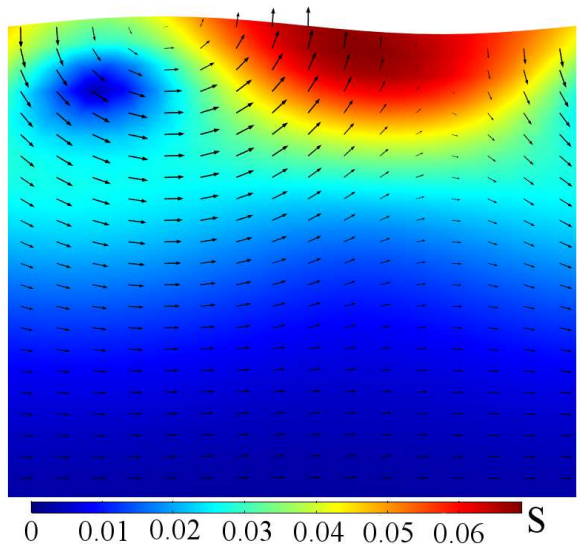}
 
 	\caption{(Color online.) Flow profile (black arrows) around the Taylor sheet superimposed with the heat map for the order parameter $S=\sqrt{Q^2_{xx}+Q^2_{xy}}$ at $a_0=0.99$, $\epsilon=bq=0.1$, $\mu=\eta=\nu/3$, $A=\nu \omega$ and $\mu_1=0$. Here the size of the simulation box is 16$\pi\times 60$ but we only show the portion of size 2$\pi \times 20$. }
 	\label{floworder}
 \end{figure}

To go beyond the restriction $\epsilon^2\ll(a_\mathrm{c}-a)^2$, we solve the 
%\rap{what is the ``momentum balance equation''? I thought it was Eq. (8). Where is the director equation in all of this?} momentum balance equation 
force balance equation $\partial_\beta\sigma_{\alpha\beta}=0$ and the director equation~\eqref{Qeq} numerically using the COMSOL Multiphysics\textregistered\ software~\cite{comsol}. We scale length by $1/q$ and time by $1/\omega$, %. We set 
and choose $\epsilon=0.1$, $\mu=\eta=\nu/3$, $A=\nu \omega$ and $\mu_1=0$. % for all the simulation studies. 
%Though it is impractical to simulate the infinitely big system but 
To approximate the infinite system, we %can 
choose the size of the simulation box much larger than the decay length  $1/q$. 
The simulation box has dimensions $32\pi$ and $60$ along the $x$ and $y$ directions, respectively, with %We use 
periodic boundary conditions along the $x$ direction. The Taylor sheet is represented by the top wall (Fig.~\ref{floworder}), which deforms and has a no-slip boundary condition. In order to ensure that the sheet is subjected to no net force along the $x$ direction, we choose the slip boundary condition $\sigma_{xy}=0$ on the bottom wall.  
More details of the numerical method are discussed in the supplementary material~\cite{supp}.

Figure~\ref{speed} shows the numerically calculated $U$ vs. $a_0$  for $a<a_\mathrm{c}$. The speed $U$ increases with $a$ monotonically, with good agreement between the simulations and theory when $a<0.9a_\mathrm{c}$. 
At $a=0.99a_\mathrm{c}$, the swimming speed is enhanced up to around 12 times the swimming speed of the Taylor case (see the inset of Fig.~\ref{speed}). We do not perform numerical studies much closer to the critical activity because the decay length  increases as  $a\to a_\mathrm{c}$, requiring a larger simulation box. In Fig. \ref{floworder}, we show the flow profile around the Taylor sheet superimposed with the heat map of the order parameter $S=\sqrt{Q^2_{xx}+Q^2_{xy}}$. The figure illustrates flow birefringence: $S$ attains its greatest values in the regions where shear is greatest.

The numerically calculated power exerted by the swimmer and power dissipated in the fluid are shown in  Fig. \ref{power}. The power exerted by the swimmer decreases with increasing activity (blue stars) whereas the rate of dissipation increases with increasing activity (red dots). When $a=0$, the power exerted by the swimmer equals the power dissipated in the fluid. However in the presence of activity, the swimmer does not work as hard, since part of the power generated by activity contributes to work on the swimmer, and part is dissipated in the fluid.

%\subsection{Conclusion}
We have studied the swimming of a model microorganism in the isotropic phase of an active nematic liquid crystal. %We model the micro-organisms as a Taylor sheet and a squirmer. The active nature of the nematic molecules effectively reduces the shear viscosity. 
As activity approaches the critical value at which the quiescent fluid is unstable to spontaneous shear flow, the swimming speed increases dramatically. An important extension of this work would be to study swimmers in the unsteady regime above the critical activity. 

\begin{acknowledgments} We thank Aparna Baskaran, Ray Goldstein,  David Henann, Oleg Lavrentovich,  and Sriram Ramaswamy for helpful discussions.
This work was supported in part by National Science Foundation Grant No. CBET-1437195 and National Science Foundation Grant MRSEC-1420382.
\end{acknowledgments}


\begin{thebibliography}{42}%
	\makeatletter
	\providecommand \@ifxundefined [1]{%
		\@ifx{#1\undefined}
	}%
	\providecommand \@ifnum [1]{%
		\ifnum #1\expandafter \@firstoftwo
		\else \expandafter \@secondoftwo
		\fi
	}%
	\providecommand \@ifx [1]{%
		\ifx #1\expandafter \@firstoftwo
		\else \expandafter \@secondoftwo
		\fi
	}%
	\providecommand \natexlab [1]{#1}%
	\providecommand \enquote  [1]{``#1''}%
	\providecommand \bibnamefont  [1]{#1}%
	\providecommand \bibfnamefont [1]{#1}%
	\providecommand \citenamefont [1]{#1}%
	\providecommand \href@noop [0]{\@secondoftwo}%
	\providecommand \href [0]{\begingroup \@sanitize@url \@href}%
	\providecommand \@href[1]{\@@startlink{#1}\@@href}%
	\providecommand \@@href[1]{\endgroup#1\@@endlink}%
	\providecommand \@sanitize@url [0]{\catcode `\\12\catcode `\$12\catcode
		`\&12\catcode `\#12\catcode `\^12\catcode `\_12\catcode `\%12\relax}%
	\providecommand \@@startlink[1]{}%
	\providecommand \@@endlink[0]{}%
	\providecommand \url  [0]{\begingroup\@sanitize@url \@url }%
	\providecommand \@url [1]{\endgroup\@href {#1}{\urlprefix }}%
	\providecommand \urlprefix  [0]{URL }%
	\providecommand \Eprint [0]{\href }%
	\providecommand \doibase [0]{http://dx.doi.org/}%
	\providecommand \selectlanguage [0]{\@gobble}%
	\providecommand \bibinfo  [0]{\@secondoftwo}%
	\providecommand \bibfield  [0]{\@secondoftwo}%
	\providecommand \translation [1]{[#1]}%
	\providecommand \BibitemOpen [0]{}%
	\providecommand \bibitemStop [0]{}%
	\providecommand \bibitemNoStop [0]{.\EOS\space}%
	\providecommand \EOS [0]{\spacefactor3000\relax}%
	\providecommand \BibitemShut  [1]{\csname bibitem#1\endcsname}%
	\let\auto@bib@innerbib\@empty
	%</preamble>
	\bibitem [{\citenamefont {Lauga}\ and\ \citenamefont
		{Powers}(2009)}]{LaugaPowers2009}%
	\BibitemOpen
	\bibfield  {author} {\bibinfo {author} {\bibfnamefont {E.}~\bibnamefont
			{Lauga}}\ and\ \bibinfo {author} {\bibfnamefont {T.~R.}\ \bibnamefont
			{Powers}},\ }\href@noop {} {\bibfield  {journal} {\bibinfo  {journal} {Rep.
				Prog. Phys.}\ }\textbf {\bibinfo {volume} {\textbf{72}}},\ \bibinfo {pages}
		{096601} (\bibinfo {year} {2009})}\BibitemShut {NoStop}%
	\bibitem [{\citenamefont {Ishiyama}\ \emph {et~al.}(2001)\citenamefont
		{Ishiyama}, \citenamefont {Sendoh}, \citenamefont {Yamazaki},\ and\
		\citenamefont {Aral}}]{Ishiyama_etal2001}%
	\BibitemOpen
	\bibfield  {author} {\bibinfo {author} {\bibfnamefont {K.}~\bibnamefont
			{Ishiyama}}, \bibinfo {author} {\bibfnamefont {M.}~\bibnamefont {Sendoh}},
		\bibinfo {author} {\bibfnamefont {A.}~\bibnamefont {Yamazaki}}, \ and\
		\bibinfo {author} {\bibfnamefont {K.~I.}\ \bibnamefont {Aral}},\ }\href@noop
	{} {\bibfield  {journal} {\bibinfo  {journal} {Sens. Actuators A}\ }\textbf
		{\bibinfo {volume} {\textbf{91}}},\ \bibinfo {pages} {141} (\bibinfo {year}
		{2001})}\BibitemShut {NoStop}%
	\bibitem [{\citenamefont {Dreyfus}\ \emph {et~al.}(2005)\citenamefont
		{Dreyfus}, \citenamefont {Baudry}, \citenamefont {Roper}, \citenamefont
		{Fermigier}, \citenamefont {Stone},\ and\ \citenamefont
		{Bibette}}]{dreyfus05_nature}%
	\BibitemOpen
	\bibfield  {author} {\bibinfo {author} {\bibfnamefont {R.}~\bibnamefont
			{Dreyfus}}, \bibinfo {author} {\bibfnamefont {J.}~\bibnamefont {Baudry}},
		\bibinfo {author} {\bibfnamefont {M.~L.}\ \bibnamefont {Roper}}, \bibinfo
		{author} {\bibfnamefont {M.}~\bibnamefont {Fermigier}}, \bibinfo {author}
		{\bibfnamefont {H.~A.}\ \bibnamefont {Stone}}, \ and\ \bibinfo {author}
		{\bibfnamefont {J.}~\bibnamefont {Bibette}},\ }\href@noop {} {\bibfield
		{journal} {\bibinfo  {journal} {Nature}\ }\textbf {\bibinfo {volume} {437}},\
		\bibinfo {pages} {862} (\bibinfo {year} {2005})}\BibitemShut {NoStop}%
	\bibitem [{\citenamefont {Ghosh}\ and\ \citenamefont
		{Fischer}(2009)}]{GhoshFischer2009}%
	\BibitemOpen
	\bibfield  {author} {\bibinfo {author} {\bibfnamefont {A.}~\bibnamefont
			{Ghosh}}\ and\ \bibinfo {author} {\bibfnamefont {P.}~\bibnamefont
			{Fischer}},\ }\href@noop {} {\bibfield  {journal} {\bibinfo  {journal} {Nano
				Lett.}\ }\textbf {\bibinfo {volume} {\textbf{9}}},\ \bibinfo {pages} {2243}
		(\bibinfo {year} {2009})}\BibitemShut {NoStop}%
	\bibitem [{\citenamefont {Guasto}\ \emph {et~al.}(2010)\citenamefont {Guasto},
		\citenamefont {Johnson},\ and\ \citenamefont {Gollub}}]{Guasto:2010dz}%
	\BibitemOpen
	\bibfield  {author} {\bibinfo {author} {\bibfnamefont {J.}~\bibnamefont
			{Guasto}}, \bibinfo {author} {\bibfnamefont {K.}~\bibnamefont {Johnson}}, \
		and\ \bibinfo {author} {\bibfnamefont {J.}~\bibnamefont {Gollub}},\
	}\href@noop {} {\bibfield  {journal} {\bibinfo  {journal} {Phys. Rev. Lett.}\
	}\textbf {\bibinfo {volume} {\textbf{105}}},\ \bibinfo {pages} {168102}
	(\bibinfo {year} {2010})}\BibitemShut {NoStop}%
\bibitem [{\citenamefont {Drescher}\ \emph {et~al.}(2011)\citenamefont
	{Drescher}, \citenamefont {Dunkel}, \citenamefont {Cisneros}, \citenamefont
	{Ganguly},\ and\ \citenamefont {Goldstein}}]{Drescheretal2011}%
\BibitemOpen
\bibfield  {author} {\bibinfo {author} {\bibfnamefont {K.}~\bibnamefont
		{Drescher}}, \bibinfo {author} {\bibfnamefont {J.}~\bibnamefont {Dunkel}},
	\bibinfo {author} {\bibfnamefont {L.~H.}\ \bibnamefont {Cisneros}}, \bibinfo
	{author} {\bibfnamefont {S.}~\bibnamefont {Ganguly}}, \ and\ \bibinfo
	{author} {\bibfnamefont {R.~E.}\ \bibnamefont {Goldstein}},\ }\href@noop {}
{\bibfield  {journal} {\bibinfo  {journal} {Proc. Natl. Acad. Sci. (USA)}\
	}\textbf {\bibinfo {volume} {\textbf{108}}},\ \bibinfo {pages} {10940}
	(\bibinfo {year} {2011})}\BibitemShut {NoStop}%
\bibitem [{\citenamefont {Goldstein}(2015)}]{Goldstein2015}%
\BibitemOpen
\bibfield  {author} {\bibinfo {author} {\bibfnamefont {R.~E.}\ \bibnamefont
		{Goldstein}},\ }\href@noop {} {\bibfield  {journal} {\bibinfo  {journal}
		{Ann. Rev. Fluid Mech.}\ }\textbf {\bibinfo {volume} {\textbf{47}}},\
	\bibinfo {pages} {343} (\bibinfo {year} {2015})}\BibitemShut {NoStop}%
\bibitem [{\citenamefont {Stone}\ and\ \citenamefont
	{Samuel}(1996)}]{stone_samuel1996}%
\BibitemOpen
\bibfield  {author} {\bibinfo {author} {\bibfnamefont {H.}~\bibnamefont
		{Stone}}\ and\ \bibinfo {author} {\bibfnamefont {A.}~\bibnamefont {Samuel}},\
}\href@noop {} {\bibfield  {journal} {\bibinfo  {journal} {Phys. Rev. Lett.}\
}\textbf {\bibinfo {volume} {{\bf 77}}},\ \bibinfo {pages} {4102} (\bibinfo
{year} {1996})}\BibitemShut {NoStop}%
\bibitem [{\citenamefont {Najafi}\ and\ \citenamefont
	{Golestanian}(2005)}]{najafi05}%
\BibitemOpen
\bibfield  {author} {\bibinfo {author} {\bibfnamefont {A.}~\bibnamefont
		{Najafi}}\ and\ \bibinfo {author} {\bibfnamefont {R.}~\bibnamefont
		{Golestanian}},\ }\href@noop {} {\bibfield  {journal} {\bibinfo  {journal}
		{J. Phys. Cond. Mat.}\ }\textbf {\bibinfo {volume} {17}},\ \bibinfo {pages}
	{S1203} (\bibinfo {year} {2005})}\BibitemShut {NoStop}%
\bibitem [{\citenamefont {Pak}\ and\ \citenamefont
	{Lauga}(2015)}]{PakLauga2015}%
\BibitemOpen
\bibfield  {author} {\bibinfo {author} {\bibfnamefont {O.~S.}\ \bibnamefont
		{Pak}}\ and\ \bibinfo {author} {\bibfnamefont {E.}~\bibnamefont {Lauga}},\
}in\ \href@noop {} {\emph {\bibinfo {booktitle} {Low-Reynolds-Number Flows:
		Fluid-Structure Interactions}}},\ \bibinfo {series} {RSC Soft Matter Series},
Vol.\ \bibinfo {volume} {\textbf{4}},\ \bibinfo {editor} {edited by\ \bibinfo
	{editor} {\bibfnamefont {E.}~\bibnamefont {Duprat}}\ and\ \bibinfo {editor}
	{\bibfnamefont {H.~A.}\ \bibnamefont {Stone}}}\ (\bibinfo  {publisher} {The
	Royal Society of Chemistry},\ \bibinfo {year} {2015})\ p.\ \bibinfo {pages}
{100}\BibitemShut {NoStop}%
\bibitem [{\citenamefont {Pozrikidis}(1992)}]{pozrikidis1992}%
\BibitemOpen
\bibfield  {author} {\bibinfo {author} {\bibfnamefont {C.}~\bibnamefont
		{Pozrikidis}},\ }\href@noop {} {\emph {\bibinfo {title} {Boundary integral
			and singularity methods for linearized viscous flow}}}\ (\bibinfo
{publisher} {Cambridge University Press},\ \bibinfo {address} {Cambridge},\
\bibinfo {year} {1992})\BibitemShut {NoStop}%
\bibitem [{\citenamefont {Cortez}(2001)}]{Cortez2001}%
\BibitemOpen
\bibfield  {author} {\bibinfo {author} {\bibfnamefont {S.}~\bibnamefont
		{Cortez}},\ }\href@noop {} {\bibfield  {journal} {\bibinfo  {journal} {{SIAM}
			J. Sci. Comput.}\ }\textbf {\bibinfo {volume} {{\bf 23}}},\ \bibinfo {pages}
	{1204} (\bibinfo {year} {2001})}\BibitemShut {NoStop}%
\bibitem [{\citenamefont {Cortez}\ \emph {et~al.}(2005)\citenamefont {Cortez},
	\citenamefont {Fauci},\ and\ \citenamefont {Medovikov}}]{cortez2}%
\BibitemOpen
\bibfield  {author} {\bibinfo {author} {\bibfnamefont {R.}~\bibnamefont
		{Cortez}}, \bibinfo {author} {\bibfnamefont {L.}~\bibnamefont {Fauci}}, \
	and\ \bibinfo {author} {\bibfnamefont {A.}~\bibnamefont {Medovikov}},\
}\href@noop {} {\bibfield  {journal} {\bibinfo  {journal} {Phys. Fluids}\
}\textbf {\bibinfo {volume} {17}},\ \bibinfo {pages} {031504} (\bibinfo
{year} {2005})}\BibitemShut {NoStop}%
\bibitem [{\citenamefont {Lauga}(2007)}]{lauga2007}%
\BibitemOpen
\bibfield  {author} {\bibinfo {author} {\bibfnamefont {E.}~\bibnamefont
		{Lauga}},\ }\href@noop {} {\bibfield  {journal} {\bibinfo  {journal} {Physics
			of Fluids}\ }\textbf {\bibinfo {volume} {19}},\ \bibinfo {pages} {083104}
	(\bibinfo {year} {2007})}\BibitemShut {NoStop}%
\bibitem [{\citenamefont {Fu}\ \emph {et~al.}(2007)\citenamefont {Fu},
	\citenamefont {Powers},\ and\ \citenamefont
	{Wolgemuth}}]{FuPowersWolgemuth2007}%
\BibitemOpen
\bibfield  {author} {\bibinfo {author} {\bibfnamefont {H.~C.}\ \bibnamefont
		{Fu}}, \bibinfo {author} {\bibfnamefont {T.~R.}\ \bibnamefont {Powers}}, \
	and\ \bibinfo {author} {\bibfnamefont {C.~W.}\ \bibnamefont {Wolgemuth}},\
}\href@noop {} {\bibfield  {journal} {\bibinfo  {journal} {Phys. Rev. Lett.}\
}\textbf {\bibinfo {volume} {{\bf 99}}},\ \bibinfo {pages} {258101} (\bibinfo
{year} {2007})}\BibitemShut {NoStop}%
\bibitem [{\citenamefont {Teran}\ \emph {et~al.}(2010)\citenamefont {Teran},
	\citenamefont {Fauci},\ and\ \citenamefont
	{Shelley}}]{TeranFauciShelley2010}%
\BibitemOpen
\bibfield  {author} {\bibinfo {author} {\bibfnamefont {J.}~\bibnamefont
		{Teran}}, \bibinfo {author} {\bibfnamefont {L.}~\bibnamefont {Fauci}}, \ and\
	\bibinfo {author} {\bibfnamefont {M.}~\bibnamefont {Shelley}},\ }\href@noop
{} {\bibfield  {journal} {\bibinfo  {journal} {Phys. Rev. Lett.}\ }\textbf
	{\bibinfo {volume} {\textbf{104}}},\ \bibinfo {pages} {038101} (\bibinfo
	{year} {2010})}\BibitemShut {NoStop}%
\bibitem [{\citenamefont {Shen}\ and\ \citenamefont
	{Arratia}(2011)}]{ShenArratia2011}%
\BibitemOpen
\bibfield  {author} {\bibinfo {author} {\bibfnamefont {X.~N.}\ \bibnamefont
		{Shen}}\ and\ \bibinfo {author} {\bibfnamefont {P.~E.}\ \bibnamefont
		{Arratia}},\ }\href@noop {} {\bibfield  {journal} {\bibinfo  {journal} {Phys.
			Rev. Lett.}\ }\textbf {\bibinfo {volume} {\textbf{106}}},\ \bibinfo {pages}
	{208101} (\bibinfo {year} {2011})}\BibitemShut {NoStop}%
\bibitem [{\citenamefont {Liu}\ \emph {et~al.}(2011)\citenamefont {Liu},
	\citenamefont {Powers},\ and\ \citenamefont {Breuer}}]{LiuPowersBreuer2011}%
\BibitemOpen
\bibfield  {author} {\bibinfo {author} {\bibfnamefont {B.}~\bibnamefont
		{Liu}}, \bibinfo {author} {\bibfnamefont {T.~R.}\ \bibnamefont {Powers}}, \
	and\ \bibinfo {author} {\bibfnamefont {K.~S.}\ \bibnamefont {Breuer}},\
}\href@noop {} {\bibfield  {journal} {\bibinfo  {journal} {Proc. Natl. Acad.
		Sci. (USA)}\ }\textbf {\bibinfo {volume} {{\bf 108}}},\ \bibinfo {pages}
{19516} (\bibinfo {year} {2011})}\BibitemShut {NoStop}%
\bibitem [{\citenamefont {{Spagnolie, ed.}}(2015)}]{SpagnolieBook2015}%
\BibitemOpen
\bibfield  {author} {\bibinfo {author} {\bibfnamefont {S.~E.}\ \bibnamefont
		{{Spagnolie, ed.}}},\ }\href@noop {} {\emph {\bibinfo {title} {Complex Fluids
			in Biological Systems}}}\ (\bibinfo  {publisher} {Springer},\ \bibinfo
{address} {New York},\ \bibinfo {year} {2015})\BibitemShut {NoStop}%
\bibitem [{\citenamefont {Zhou}\ \emph {et~al.}(2014)\citenamefont {Zhou},
	\citenamefont {Sokolov}, \citenamefont {Lavrentovich},\ and\ \citenamefont
	{Aranson}}]{Zhou_etal2013}%
\BibitemOpen
\bibfield  {author} {\bibinfo {author} {\bibfnamefont {S.}~\bibnamefont
		{Zhou}}, \bibinfo {author} {\bibfnamefont {A.}~\bibnamefont {Sokolov}},
	\bibinfo {author} {\bibfnamefont {O.~D.}\ \bibnamefont {Lavrentovich}}, \
	and\ \bibinfo {author} {\bibfnamefont {I.~S.}\ \bibnamefont {Aranson}},\
}\href@noop {} {\bibfield  {journal} {\bibinfo  {journal} {Proc. Natl. Acad.
		Sci. USA}\ }\textbf {\bibinfo {volume} {\textbf{111}}},\ \bibinfo {pages}
{1265} (\bibinfo {year} {2014})}\BibitemShut {NoStop}%
\bibitem [{\citenamefont {Mushenheim}\ \emph {et~al.}(2014)\citenamefont
	{Mushenheim}, \citenamefont {Trivedi}, \citenamefont {Tuson}, \citenamefont
	{Weibel},\ and\ \citenamefont {Abbott}}]{MushenheimEtAl2013}%
\BibitemOpen
\bibfield  {author} {\bibinfo {author} {\bibfnamefont {P.~C.}\ \bibnamefont
		{Mushenheim}}, \bibinfo {author} {\bibfnamefont {R.~R.}\ \bibnamefont
		{Trivedi}}, \bibinfo {author} {\bibfnamefont {H.~H.}\ \bibnamefont {Tuson}},
	\bibinfo {author} {\bibfnamefont {D.~B.}\ \bibnamefont {Weibel}}, \ and\
	\bibinfo {author} {\bibfnamefont {N.~L.}\ \bibnamefont {Abbott}},\
}\href@noop {} {\bibfield  {journal} {\bibinfo  {journal} {Soft Matter}\
}\textbf {\bibinfo {volume} {\textbf{10}}},\ \bibinfo {pages} {88} (\bibinfo
{year} {2014})}\BibitemShut {NoStop}%
\bibitem [{\citenamefont {Krieger}\ \emph {et~al.}(2014)\citenamefont
	{Krieger}, \citenamefont {Spagnolie},\ and\ \citenamefont
	{Powers}}]{KriegerSpagnoliePowers2014}%
\BibitemOpen
\bibfield  {author} {\bibinfo {author} {\bibfnamefont {M.~S.}\ \bibnamefont
		{Krieger}}, \bibinfo {author} {\bibfnamefont {S.~E.}\ \bibnamefont
		{Spagnolie}}, \ and\ \bibinfo {author} {\bibfnamefont {T.~R.}\ \bibnamefont
		{Powers}},\ }\href@noop {} {\bibfield  {journal} {\bibinfo  {journal} {Phys.
			Rev. E}\ }\textbf {\bibinfo {volume} {\textbf{90}}},\ \bibinfo {pages}
	{052503} (\bibinfo {year} {2014})}\BibitemShut {NoStop}%
\bibitem [{\citenamefont {Marchetti}\ \emph {et~al.}(2013)\citenamefont
	{Marchetti}, \citenamefont {Joanny}, \citenamefont {Ramaswamy}, \citenamefont
	{Liverpool}, \citenamefont {Prost}, \citenamefont {Rao},\ and\ \citenamefont
	{Simha}}]{srimrmp}%
\BibitemOpen
\bibfield  {author} {\bibinfo {author} {\bibfnamefont {M.~C.}\ \bibnamefont
		{Marchetti}}, \bibinfo {author} {\bibfnamefont {J.~F.}\ \bibnamefont
		{Joanny}}, \bibinfo {author} {\bibfnamefont {S.}~\bibnamefont {Ramaswamy}},
	\bibinfo {author} {\bibfnamefont {T.~B.}\ \bibnamefont {Liverpool}}, \bibinfo
	{author} {\bibfnamefont {J.}~\bibnamefont {Prost}}, \bibinfo {author}
	{\bibfnamefont {M.}~\bibnamefont {Rao}}, \ and\ \bibinfo {author}
	{\bibfnamefont {R.~A.}\ \bibnamefont {Simha}},\ }\href {\doibase
	10.1103/RevModPhys.85.1143} {\bibfield  {journal} {\bibinfo  {journal} {Rev.
			Mod. Phys.}\ }\textbf {\bibinfo {volume} {\textbf{85}}},\ \bibinfo {pages}
	{1143} (\bibinfo {year} {2013})}\BibitemShut {NoStop}%
\bibitem [{\citenamefont {Schaller}\ \emph {et~al.}(2010)\citenamefont
	{Schaller}, \citenamefont {Weber}, \citenamefont {Semmrich}, \citenamefont
	{Frey},\ and\ \citenamefont {Bausch}}]{Schaller2010}%
\BibitemOpen
\bibfield  {author} {\bibinfo {author} {\bibfnamefont {V.}~\bibnamefont
		{Schaller}}, \bibinfo {author} {\bibfnamefont {C.}~\bibnamefont {Weber}},
	\bibinfo {author} {\bibfnamefont {C.}~\bibnamefont {Semmrich}}, \bibinfo
	{author} {\bibfnamefont {E.}~\bibnamefont {Frey}}, \ and\ \bibinfo {author}
	{\bibfnamefont {A.~R.}\ \bibnamefont {Bausch}},\ }\href {\doibase
	10.1038/nature09312} {\bibfield  {journal} {\bibinfo  {journal} {Nature}\
	}\textbf {\bibinfo {volume} {467}},\ \bibinfo {pages} {73} (\bibinfo {year}
	{2010})}\BibitemShut {NoStop}%
\bibitem [{\citenamefont {Sanchez}\ \emph {et~al.}(2012)\citenamefont
	{Sanchez}, \citenamefont {Chen}, \citenamefont {DeCamp}, \citenamefont
	{Heymann},\ and\ \citenamefont {Dogic}}]{Sanchez2012}%
\BibitemOpen
\bibfield  {author} {\bibinfo {author} {\bibfnamefont {T.}~\bibnamefont
		{Sanchez}}, \bibinfo {author} {\bibfnamefont {D.~T.~N.}\ \bibnamefont
		{Chen}}, \bibinfo {author} {\bibfnamefont {S.~J.}\ \bibnamefont {DeCamp}},
	\bibinfo {author} {\bibfnamefont {M.}~\bibnamefont {Heymann}}, \ and\
	\bibinfo {author} {\bibfnamefont {Z.}~\bibnamefont {Dogic}},\ }\href
{\doibase 10.1038/nature11591} {\bibfield  {journal} {\bibinfo  {journal}
		{Nature}\ }\textbf {\bibinfo {volume} {491}},\ \bibinfo {pages} {431}
	(\bibinfo {year} {2012})}\BibitemShut {NoStop}%
\bibitem [{\citenamefont {Taylor}(1951)}]{taylor1951}%
\BibitemOpen
\bibfield  {author} {\bibinfo {author} {\bibfnamefont {G.}~\bibnamefont
		{Taylor}},\ }\href {\doibase 10.1098/rspa.1951.0218} {\bibfield  {journal}
	{\bibinfo  {journal} {Proceedings of the Royal Society of London A:
			Mathematical, Physical and Engineering Sciences}\ }\textbf {\bibinfo {volume}
		{209}},\ \bibinfo {pages} {447} (\bibinfo {year} {1951})}\BibitemShut
{NoStop}%
\bibitem [{\citenamefont {Giomi}\ \emph {et~al.}(2012)\citenamefont {Giomi},
	\citenamefont {Mahadevan}, \citenamefont {Chakraborty},\ and\ \citenamefont
	{Hagan}}]{Giomi_etal2012}%
\BibitemOpen
\bibfield  {author} {\bibinfo {author} {\bibfnamefont {L.}~\bibnamefont
		{Giomi}}, \bibinfo {author} {\bibfnamefont {L.}~\bibnamefont {Mahadevan}},
	\bibinfo {author} {\bibfnamefont {B.}~\bibnamefont {Chakraborty}}, \ and\
	\bibinfo {author} {\bibfnamefont {M.~F.}\ \bibnamefont {Hagan}},\ }\href@noop
{} {\bibfield  {journal} {\bibinfo  {journal} {Nonlinearity}\ }\textbf
	{\bibinfo {volume} {\textbf{25}}},\ \bibinfo {pages} {2245} (\bibinfo {year}
	{2012})}\BibitemShut {NoStop}%
\bibitem [{\citenamefont {{d}e Gennes}(1969)}]{DeGennes1969}%
\BibitemOpen
\bibfield  {author} {\bibinfo {author} {\bibfnamefont {P.~G.}\ \bibnamefont
		{{d}e Gennes}},\ }\href {\doibase
	http://dx.doi.org/10.1016/0375-9601(69)90240-0} {\bibfield  {journal}
	{\bibinfo  {journal} {Physics Letters A}\ }\textbf {\bibinfo {volume} {30}},\
	\bibinfo {pages} {454 } (\bibinfo {year} {1969})}\BibitemShut {NoStop}%
\bibitem [{\citenamefont {de~Gennes}\ and\ \citenamefont
	{Prost}(1993)}]{degennesbook}%
\BibitemOpen
\bibfield  {author} {\bibinfo {author} {\bibfnamefont {P.~G.}\ \bibnamefont
		{de~Gennes}}\ and\ \bibinfo {author} {\bibfnamefont {J.}~\bibnamefont
		{Prost}},\ }\href@noop {} {\emph {\bibinfo {title} {The physics of liquid
			crystals}}}\ (\bibinfo  {publisher} {Clarendon Press Oxford University
	Press},\ \bibinfo {address} {Oxford New York},\ \bibinfo {year}
{1993})\BibitemShut {NoStop}%
\bibitem [{\citenamefont {de~Gennes}(1971)}]{deGennes1971}%
\BibitemOpen
\bibfield  {author} {\bibinfo {author} {\bibfnamefont {P.~G.}\ \bibnamefont
		{de~Gennes}},\ }\href {\doibase 10.1080/15421407108082773} {\bibfield
	{journal} {\bibinfo  {journal} {Molecular Crystals and Liquid Crystals}\
	}\textbf {\bibinfo {volume} {12}},\ \bibinfo {pages} {193} (\bibinfo {year}
	{1971})}\BibitemShut {NoStop}%
\bibitem [{\citenamefont {Hatwalne}\ \emph {et~al.}(2004)\citenamefont
	{Hatwalne}, \citenamefont {Ramaswamy}, \citenamefont {Rao},\ and\
	\citenamefont {Simha}}]{Hatwalne2004}%
\BibitemOpen
\bibfield  {author} {\bibinfo {author} {\bibfnamefont {Y.}~\bibnamefont
		{Hatwalne}}, \bibinfo {author} {\bibfnamefont {S.}~\bibnamefont {Ramaswamy}},
	\bibinfo {author} {\bibfnamefont {M.}~\bibnamefont {Rao}}, \ and\ \bibinfo
	{author} {\bibfnamefont {R.~A.}\ \bibnamefont {Simha}},\ }\href {\doibase
	10.1103/PhysRevLett.92.118101} {\bibfield  {journal} {\bibinfo  {journal}
		{Phys. Rev. Lett.}\ }\textbf {\bibinfo {volume} {92}},\ \bibinfo {pages}
	{118101} (\bibinfo {year} {2004})}\BibitemShut {NoStop}%
\bibitem [{\citenamefont {Saintillan}\ and\ \citenamefont
	{Shelley}(2008)}]{SaintillanShelley2008}%
\BibitemOpen
\bibfield  {author} {\bibinfo {author} {\bibfnamefont {D.}~\bibnamefont
		{Saintillan}}\ and\ \bibinfo {author} {\bibfnamefont {M.~J.}\ \bibnamefont
		{Shelley}},\ }\href@noop {} {\bibfield  {journal} {\bibinfo  {journal} {Phys.
			Rev. Lett.}\ }\textbf {\bibinfo {volume} {\textbf{100}}},\ \bibinfo {pages}
	{178103} (\bibinfo {year} {2008})}\BibitemShut {NoStop}%
\bibitem [{\citenamefont {Woodhouse}\ and\ \citenamefont
	{Goldstein}(2012)}]{WoodhouseGoldstein2012}%
\BibitemOpen
\bibfield  {author} {\bibinfo {author} {\bibfnamefont {F.~G.}\ \bibnamefont
		{Woodhouse}}\ and\ \bibinfo {author} {\bibfnamefont {R.~E.}\ \bibnamefont
		{Goldstein}},\ }\href@noop {} {\bibfield  {journal} {\bibinfo  {journal}
		{Phys. Rev. Lett.}\ }\textbf {\bibinfo {volume} {\textbf{109}}},\ \bibinfo
	{pages} {168105} (\bibinfo {year} {2012})}\BibitemShut {NoStop}%
\bibitem [{\citenamefont {Cates}\ \emph {et~al.}(2008)\citenamefont {Cates},
	\citenamefont {Fielding}, \citenamefont {Marenduzzo}, \citenamefont
	{Orlandini},\ and\ \citenamefont {Yeomans}}]{Cates_etal2008}%
\BibitemOpen
\bibfield  {author} {\bibinfo {author} {\bibfnamefont {M.~E.}\ \bibnamefont
		{Cates}}, \bibinfo {author} {\bibfnamefont {S.~M.}\ \bibnamefont {Fielding}},
	\bibinfo {author} {\bibfnamefont {D.}~\bibnamefont {Marenduzzo}}, \bibinfo
	{author} {\bibfnamefont {E.}~\bibnamefont {Orlandini}}, \ and\ \bibinfo
	{author} {\bibfnamefont {J.~M.}\ \bibnamefont {Yeomans}},\ }\href@noop {}
{\bibfield  {journal} {\bibinfo  {journal} {Phys. Rev. Lett.}\ }\textbf
	{\bibinfo {volume} {\textbf{101}}},\ \bibinfo {pages} {068102} (\bibinfo
	{year} {2008})}\BibitemShut {NoStop}%
\bibitem [{\citenamefont {Sokolov}\ and\ \citenamefont
	{Aranson}(2009)}]{Sokolov2009}%
\BibitemOpen
\bibfield  {author} {\bibinfo {author} {\bibfnamefont {A.}~\bibnamefont
		{Sokolov}}\ and\ \bibinfo {author} {\bibfnamefont {I.~S.}\ \bibnamefont
		{Aranson}},\ }\href {\doibase 10.1103/PhysRevLett.103.148101} {\bibfield
	{journal} {\bibinfo  {journal} {Phys. Rev. Lett.}\ }\textbf {\bibinfo
		{volume} {103}},\ \bibinfo {pages} {148101} (\bibinfo {year}
	{2009})}\BibitemShut {NoStop}%
\bibitem [{\citenamefont {Giomi}\ \emph {et~al.}(2010)\citenamefont {Giomi},
	\citenamefont {Liverpool},\ and\ \citenamefont
	{Marchetti}}]{GiomiLiverpoolMarchetti2010}%
\BibitemOpen
\bibfield  {author} {\bibinfo {author} {\bibfnamefont {L.}~\bibnamefont
		{Giomi}}, \bibinfo {author} {\bibfnamefont {T.~B.}\ \bibnamefont
		{Liverpool}}, \ and\ \bibinfo {author} {\bibfnamefont {M.~C.}\ \bibnamefont
		{Marchetti}},\ }\href@noop {} {\bibfield  {journal} {\bibinfo  {journal}
		{Phys. Rev. E}\ }\textbf {\bibinfo {volume} {\textbf{81}}},\ \bibinfo {pages}
	{051908} (\bibinfo {year} {2010})}\BibitemShut {NoStop}%
\bibitem [{\citenamefont {Gachelin}\ \emph {et~al.}(2013)\citenamefont
	{Gachelin}, \citenamefont {Mi\~no}, \citenamefont {Berthet}, \citenamefont
	{Lindner}, \citenamefont {Rousselet},\ and\ \citenamefont
	{Cl\'ement}}]{Gachelin2013}%
\BibitemOpen
\bibfield  {author} {\bibinfo {author} {\bibfnamefont {J.}~\bibnamefont
		{Gachelin}}, \bibinfo {author} {\bibfnamefont {G.}~\bibnamefont {Mi\~no}},
	\bibinfo {author} {\bibfnamefont {H.}~\bibnamefont {Berthet}}, \bibinfo
	{author} {\bibfnamefont {A.}~\bibnamefont {Lindner}}, \bibinfo {author}
	{\bibfnamefont {A.}~\bibnamefont {Rousselet}}, \ and\ \bibinfo {author}
	{\bibfnamefont {E.}~\bibnamefont {Cl\'ement}},\ }\href {\doibase
	10.1103/PhysRevLett.110.268103} {\bibfield  {journal} {\bibinfo  {journal}
		{Phys. Rev. Lett.}\ }\textbf {\bibinfo {volume} {110}},\ \bibinfo {pages}
	{268103} (\bibinfo {year} {2013})}\BibitemShut {NoStop}%
\bibitem [{\citenamefont {L\'opez}\ \emph {et~al.}(2015)\citenamefont
	{L\'opez}, \citenamefont {Gachelin}, \citenamefont {Douarche}, \citenamefont
	{Auradou},\ and\ \citenamefont {Cl\'ement}}]{Lopez2015}%
\BibitemOpen
\bibfield  {author} {\bibinfo {author} {\bibfnamefont {H.~M.}\ \bibnamefont
		{L\'opez}}, \bibinfo {author} {\bibfnamefont {J.}~\bibnamefont {Gachelin}},
	\bibinfo {author} {\bibfnamefont {C.}~\bibnamefont {Douarche}}, \bibinfo
	{author} {\bibfnamefont {H.}~\bibnamefont {Auradou}}, \ and\ \bibinfo
	{author} {\bibfnamefont {E.}~\bibnamefont {Cl\'ement}},\ }\href {\doibase
	10.1103/PhysRevLett.115.028301} {\bibfield  {journal} {\bibinfo  {journal}
		{Phys. Rev. Lett.}\ }\textbf {\bibinfo {volume} {115}},\ \bibinfo {pages}
	{028301} (\bibinfo {year} {2015})}\BibitemShut {NoStop}%
\bibitem [{\citenamefont {J\"ulicher}\ \emph {et~al.}(2018)\citenamefont
	{J\"ulicher}, \citenamefont {Grill},\ and\ \citenamefont
	{Salbreux}}]{JulicherGrillSalbreaux2018}%
\BibitemOpen
\bibfield  {author} {\bibinfo {author} {\bibfnamefont {F.}~\bibnamefont
		{J\"ulicher}}, \bibinfo {author} {\bibfnamefont {S.~W.}\ \bibnamefont
		{Grill}}, \ and\ \bibinfo {author} {\bibfnamefont {G.}~\bibnamefont
		{Salbreux}},\ }\href@noop {} {\bibfield  {journal} {\bibinfo  {journal} {Rep.
			Prog. Phys.}\ }\textbf {\bibinfo {volume} {\textbf{81}}},\ \bibinfo {pages}
	{076601} (\bibinfo {year} {2018})}\BibitemShut {NoStop}%
\bibitem [{\citenamefont {Onsager}(1931)}]{onsager}%
\BibitemOpen
\bibfield  {author} {\bibinfo {author} {\bibfnamefont {L.}~\bibnamefont
		{Onsager}},\ }\href {\doibase 10.1103/PhysRev.37.405} {\bibfield  {journal}
	{\bibinfo  {journal} {Phys. Rev.}\ }\textbf {\bibinfo {volume} {37}},\
	\bibinfo {pages} {405} (\bibinfo {year} {1931})}\BibitemShut {NoStop}%
\bibitem [{sup()}]{supp}%
\BibitemOpen
\href@noop {} {\bibinfo  {journal} {See Supplemental material at [] for more
		detail.}\ }\BibitemShut {NoStop}%
\bibitem [{com()}]{comsol}%
\BibitemOpen
\bibfield  {journal} {  }\href@noop {} {\bibinfo  {journal} {COMSOL
		Multiphysics® v. 5.2. www.comsol.com. COMSOL AB, Stockholm, Sweden}\
}\BibitemShut {NoStop}%
\end{thebibliography}
\end{document}

\begin{center}
\Large{\textbf{\underline{Supplemental Material}}}
\end{center}

\title{Enhancement of microorganism swimming speed  in active matter}

\author{Harsh Soni$^1$}
\author{Robert A. Pelcovits$^2$}
\author{Thomas R. Powers$^{1,2}$}

\affiliation{$^1$School of Engineering, Brown University, Providence, RI 02912 USA}
\affiliation{$^2$Department of Physics, Brown University, Providence, RI 02012 USA}
\maketitle
\section{Stability analysis}
Our equations of motion are given by
\begin{eqnarray}
-\partial_\alpha p+ 2\eta\partial_\alpha e_{\beta\alpha}+2 (\mu+\mu_1) \partial_\alpha R_{\beta\alpha}+a\partial_\alpha Q_{\beta\alpha}&=&0\label{eqfb1}\\
AQ_{\alpha\beta}+2\mu e_{\alpha\beta}+\nu R_{\alpha\beta}&=&0.\label{Qeqn1}
\end{eqnarray}
corresponding respectively to %Eq.~(8) (using Eqs.~(6) and (7)) 
$\partial_\beta\sigma_{\alpha\beta}=0$ with the Ericksen stress $\sigma^\mathrm{e}_{\alpha\beta}=A Q_{\gamma\delta}Q_{\gamma\delta}\delta_{\alpha\beta}$ absorbed into pressure,
and Eq.~(5) (using $\Phi_{\alpha\beta}=-AQ_{\alpha\beta}$) of the main paper.

Eliminating $R_{\alpha\beta}$ from  Eq. \eqref{eqfb1} using Eq. \eqref{Qeqn1} we find
\begin{equation}\label{stbeq1}
-\bm{\nabla} p+ 2\left( \eta-\dfrac{2 \mu (\mu+\mu_1)}{\nu}\right) \bm{\nabla}\cdot \bm{e}+\left( a-\dfrac{2 (\mu+\mu_1) A}{\nu}\right)  \bm{\nabla}\cdot\bm{Q}=0.
\end{equation}	
To leading order Eq. \eqref{Qeqn1} becomes [using Eq.~(3) of the main paper]
\begin{equation}\label{stbeq2}
\partial_t \bm{Q}=-\dfrac{1}{\nu}(A\bm{Q}+2\mu \bm{e}).
\end{equation}
Applying the operations $\bm{\nabla}\cdot(\bm{\nabla}\cdot\quad)$  and $\bm{\nabla}\times(\bm{\nabla}\cdot\quad)$ to %the above equation, we find respectively
Eq.~(\ref{stbeq2}), we find
\begin{eqnarray}
\partial_t X&=&-\dfrac{A}{\nu} X\label{stbeq3}\\
\partial_t \bm{Y}&=&-\dfrac{A}{\nu} \bm{Y}-\dfrac{2\mu}{\nu} \bm{\nabla}\times(\bm{\nabla}\cdot\bm{e}),\label{stbeq4}
\end{eqnarray}
where $X=\bm{\nabla}\cdot(\bm{\nabla}\cdot\bm{Q})$ and $\bm{Y}=\bm{\nabla}\times(\bm{\nabla}\cdot\bm{Q})$. Taking the curl of Eq. \eqref{stbeq1} yields
\begin{equation}\label{stbeq5}
2\left( \eta-\dfrac{2 \mu (\mu+\mu_1)}{\nu}\right) \bm{\nabla}\times(\bm{\nabla}\cdot\bm{e})+\left( a-\dfrac{2 (\mu+\mu_1) A}{\nu}\right) \bm{Y}=0.
\end{equation}
Eliminating $ \bm{\nabla}\times(\bm{\nabla}\cdot\bm{e})$ from Eq. \eqref{stbeq4} using Eq. \eqref{stbeq5}, we obtain
\begin{equation}\label{stbeq6}
\partial_t \bm{Y}=-\frac{A \eta -a \mu }{\eta  \nu-2 \mu (\mu+\mu_1)} \bm{Y}.
\end{equation}
Eqs.~\eqref{stbeq3} and \eqref{stbeq6} yield the growth rates shown in Eqs.~(12) and (13) of the main paper (using the definitions of the effective viscosities given by Eqs.~(6) and (7) of the main paper) and indicate that the mode describing the dynamics of  $\bm{\nabla}\cdot(\bm{\nabla}\cdot\bm{Q})$ always decays with time, whereas the mode describing the dynamics of $\bm{\nabla}\times(\bm{\nabla}\cdot\bm{Q})$ can grow with time if the factor $-(A \eta -a \mu )/(\eta  \nu-2 \mu (\mu+\mu_1))$ is positive.

\section{Next order term in Swimming speed $U$}

To find the form of the correction in $\epsilon$ to the leading order expression for the swimming speed shown in Eq.~(19) of the main paper, we consider the general form of the solution to the equations of motion \eqref{eqfb1} and \eqref{Qeqn1} of the previous section with the boundary conditions specified in the main paper. The solution has the following form:
\begin{eqnarray}
\phi=&&\sum_{n=1,3,5...}^{\infty}\sum_{m=1,3,5...}^{n} \epsilon^n\phi^{(n,m)}(y)\exp(m i(qx-\omega t))\nonumber\\&&+\sum_{n=0,2,4...}^{\infty}\sum_{m=0,2,4...}^{n} \epsilon^n\phi^{(n,m)}(y)\exp(m i(qx-\omega t)),
\end{eqnarray}
where $\phi$ denotes $\psi$, $Q_{\alpha \beta}$, $p$, and the nonlinear part of $R_{\alpha\beta}$, $N_{\alpha\beta}=\mathbf{v}\cdot\boldsymbol{\nabla}Q_{\alpha\beta}+\omega_{\alpha\gamma}Q_{\gamma\beta}-Q_{\alpha\gamma}\omega_{\gamma\beta}$. 
%to the $n^{th}$ order in $\epsilon$ is $\sum_{m=1,3,5..n}R^{(n,m)}_{\alpha\beta}(y)\exp(m i(qx-\omega t))$ for odd $n$ and $\sum_{m=0,2,4..n}R^{(n,m)}_{\alpha\beta}(y)\exp(m i(qx-\omega t))$ for even $n$. 
Inserting these forms into Eqs. \eqref{eqfb1} and \eqref{Qeqn1} we obtain, after some straightforward calculation,
\begin{subequations}
	\begin{eqnarray}
\psi^{(n,m)}(y)&=&(C_1+C_2  y)\exp(-mqy)+\dfrac{2 A\mu_\text{eff}}{A \eta_\text{eff}+i m  \omega (2 \mu\mu_\text{eff}-\eta_\text{eff}\nu)}M^{(n,m)}(y),\label{form1}\\
Q_{xx}^{(n,m)}(y)&=& \dfrac{-1}{A-im\nu\omega}\Bigg(\nu N^{(n,m)}_{xx}(y)+2 i m q \mu\Bigg((C_2(1-mqy)-C_1m q)\exp(-m q y)\Bigg.\Bigg.\nonumber\\&+&\Bigg.\Bigg.\dfrac{2 A\mu_\text{eff}}{A \eta_\text{eff}+i m  \omega (2 \mu\mu_\text{eff}-\eta_\text{eff}\nu)}DM^{(n,m)}(y)\Bigg)\Bigg),\label{form2}\\
Q_{xy}^{(n,m)}(y)&=& \dfrac{-1}{A-im\nu\omega}\Bigg(\nu N^{(n,m)}_{xy}(y)+\Bigg(2 m q \mu(C_2(-1+mqy)+C_1m q)\exp(-m q y)\Bigg.\Bigg.\nonumber\\&+&\Bigg.\Bigg.\dfrac{2 A\mu_\text{eff}}{A \eta_\text{eff}+i m  \omega (2 \mu\mu_\text{eff}-\eta_\text{eff}\nu)}(D^2+m^2q^2)M^{(n,m)}(y)\Bigg)\Bigg),\label{form3}
\end{eqnarray}\label{form}
\end{subequations}
where $C_1$ and $C_2$ are constants which depend on $n$ and $m$, and 
\begin{equation}
M^{(n,m)}(y)=\dfrac{-1}{(D-mq)^2(D+mq)^2}(m^2 q^2N^{(n,m)}_{xy}(y)+2 i m q DN^{(n,m)}_{xx}(y)+D^2N^{(n,m)}_{xy}(y)).
\end{equation}
Here $D\equiv\dfrac{d}{dy}$ and the operator $1/(D\pm m q)$ is defined as
\begin{equation}
\dfrac{1}{D\pm m q}H(y)=\exp(\mp mqy)\int \exp(\pm mqy)H(y)dy, 
\end{equation}
where $H(y)$ is an arbitrary function.
Note that the $n^{th}$ order nonlinear term $N^{(n,m)}_{\alpha \beta}$ is obtained from the sum of the products of lower order terms linear in $\psi^{(k,m)}(y)$, $Q_{\alpha\beta}^{(k,m)}(y)$ ($k<n$) and their derivatives. 
The first order terms in $N_{\alpha \beta}$ are zero. Therefore $N^{(1,1)}_{\alpha \beta}(y)=M^{(1,1)}(y)=0$, and $\psi^{(1,1)}(y)$ and $ Q_{\alpha\beta}^{(1,1)}(y)$ are independent of  $(a_c - a)$. 
The second order terms in $N_{\alpha \beta}$ are second order in $\psi^{(1,1)}(y)$, $ Q_{\alpha\beta}^{(1,1)}(y)$ and their derivatives. Since $\psi^{(1,1)}(y)$ and $ Q_{\alpha\beta}^{(1,1)}(y)$ don't depend on $(a_c-a)$, $N^{(2,0)}_{\alpha \beta}(y)$, $M^{(2,0)}(y)$, $N^{(2,2)}_{\alpha \beta}(y)$ and $M^{(2,2)}(y)$ are also independent of  $a$. Therefore, we can see from the Eq. \eqref{form} (recalling $\eta_\text{eff}=\mu(a_c-a)/A$) that $Q_{xx}^{(2,0)}(y)$ is independent of $(a_c-a)$, and $\psi^{(2,2)}(y)$, $Q_{xx}^{(2,2)}(y)$ and  $Q_{xy}^{(2,2)}(y)$ approaching a finite limit and 
\begin{equation}
\psi^{(2,0)}(y), Q_{xy}^{(2,0)}(y)\to\dfrac{1}{(a_c-a)^1}\qquad\text{as}\qquad a\to a_c.\label{goes3}
\end{equation}
The third order terms in $N_{\alpha \beta}$ arise due to the products of $d\phi^{(1,m)}(y)/dy^k$ and $d\phi^{(2,m)}(y)/dy^k$, where $k=0,1,2$ and $\phi$ stands for $\psi$ and $Q_{\alpha\beta}$. Thus,  from \eqref{goes3},
\begin{equation}
N^{(3,1)}_{\alpha \beta}(y),M^{(3,1)}(y),N^{(3,3)}_{\alpha \beta}(y),M^{(3,3)}(y)\to\dfrac{1}{(a_c-a)^1}\qquad\text{as}\qquad a\to a_c.
\end{equation}
Hence
\begin{eqnarray}
\psi^{(3,1)}(y), Q_{\alpha\beta}^{(3,1)}(y),\psi^{(3,3)}(y), Q_{\alpha\beta}^{(3,3)}(y)\to\dfrac{1}{(a_c-a)^1}&&\qquad\text{as}\qquad a\to a_c.\label{goes4}
\end{eqnarray}
The fourth order term in $N_{\alpha \beta}$ are the linear combinations of $d\phi^{(n_1,m)}(y)/dy^kd\phi^{(n_2,m)}(y)/dy^k$, where $k=0,1,2$, $n_1+n_2=4$ and $\phi$ stands for $\psi$ and $Q_{\alpha\beta}$. Thus, from \eqref{goes3} and \eqref{goes4},
\begin{eqnarray}
N^{(4,0)}_{\alpha \beta}(y),&&M^{(4,0)}(y),N^{(4,2)}_{\alpha \beta}(y),\nonumber\\&&M^{(4,2)}(y),N^{(4,4)}_{\alpha \beta}(y),M^{(4,4)}(y)\to\dfrac{1}{(a_c-a)^2}\qquad\text{as}\qquad a\to a_c.\label{goes5}
\end{eqnarray}
Hence
\begin{eqnarray}
\qquad\qquad\qquad\qquad\qquad\qquad\psi^{(4,0)}(y)\to\dfrac{1}{(a_c-a)^3}&&\qquad\text{as}\qquad a\to a_c.
\end{eqnarray}
 Above analysis suggests that $\psi$ and $Q_{\alpha\beta}$ can be written in terms of the inverse power series of $a_c-a$, starting with the terms independent of that quantity.

Since we have $\epsilon\to-\epsilon$ symmetry, the next order correction to $U$ will be $\epsilon^4\partial_y\psi^{(4,0)}(y)$,which goes as  $\epsilon^4/(a_c-a)^{3}$ as $a\to a_c$.  Similarly, the next order term in the power dissipated in the fluid $P_\text{F}$ also goes as $\epsilon^4/(a_c-a)^{2}$. However, the power supplied by the swimmer $P_\text{S}$ goes as $\epsilon^4/(a_c-a)$ because the fluid velocity at the Taylor sheet does not depend on $a_c-a$. 
\section{Simulation details}
\begin{figure}[h]
\includegraphics[width=0.90\textwidth]{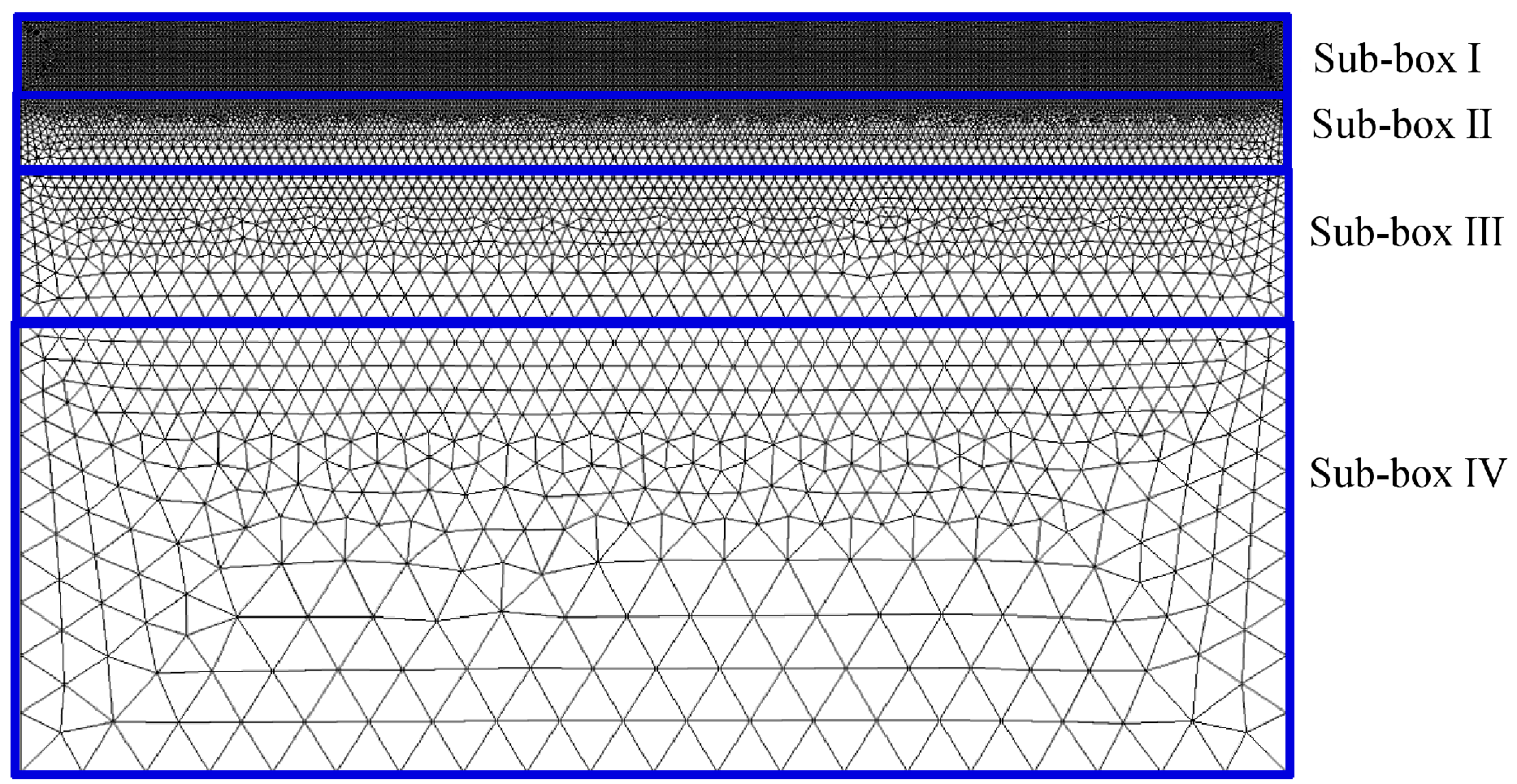}
	%\centering
\caption{(Color online.) The rectangular simulation box of size 32$\pi\times16$  is divided into four subboxes. Each subbox is further divided into triangular grids.}
 \label{box}
\end{figure}

The typical wavelength of the perturbations near the Taylor sheet is comparable to the wavelength of the sheet ($2\pi$ in dimensionless units). Thus we need to divide the region close to the sheet into grids of size much smaller than $2\pi$. Far away from the sheet, the typical wavelength of the perturbations is of the order of the system size along the $x$ direction, which is $32\pi$ in dimensionless units.  This region far from the sheet can thus be divided into larger grids. To smooth this variation in the grid size from the region near the sheet to the region near the bounding wall, we divided the simulation box into 4 subboxes along the $y$ direction, the direction perpendicular to the flat Taylor sheet (see Fig.~\ref{box}). 
 Each subbox is divided into triangular grids of different sizes, determined automatically by COMSOL after we input the maximum grid size. The sizes of the subboxes, the number of triangular grids in each subbox and the maximum size of the grids (i.e., the length of the sides of the triangles) are shown in the following table:
 
\begin{tabular}{ |p{3cm}||p{4cm}|p{3cm}|p{4.5cm}|  }
	\hline
%	\multicolumn{4}{|c|}{} \\
%	\hline
& Size of the subbox& No. of grids& Maximum grid size\\
	\hline
	Subbox I   & 32$\pi$ x 6    &15382&   0.314\\
	Subbox II& 32$\pi$ x  6  & 6736   &1.26\\
	Subbox III &32$\pi$ x 12 & 1976&  2.51\\
	Subbox IV    &32$\pi$   36 & 944&  5.03\\
	\hline
\end{tabular}\\

The Taylor sheet is represented by the top deformable wall of the simulation box; subbox I is adjacent to the sheet and subbox IV is at the bottom of the simulation box. Since small wavelength perturbations are important close to the Taylor sheet, subboxes I and II are divided into numerous small grids compared to subboxes III and IV. 

%\section{first order flow and ordering profile}
%\begin{figure}[h]
%   \centering
%   \begin{subfigure}[b]{0.45\textwidth}
%   	\includegraphics[width=\textwidth]{flow.eps}
%   	\caption{Flow profile}
%   \end{subfigure}
%   \begin{subfigure}[b]{0.45\textwidth}
%   	\includegraphics[width=\textwidth]{orientation.eps}
%   	\caption{Orientation profile }
%   \end{subfigure}
%   \caption{First order flow and ordering profile around the Taylor sheet in the frame moving with the sheet. This figure shows the that birefringence effect leads to the local ordering in the system.}\label{fig:animals}
%\end{figure}
\section{Swimming speed calculation for the squirmer}
We consider a cylindrical squirmer~\cite{blake1971} of radius $R$ subject to surface waves defined in polar coordinates by
\begin{eqnarray}
r&=&R[1+\epsilon\Delta r],\\
\phi&=&\theta+\epsilon \Delta \theta,
\end{eqnarray}
where 
\begin{eqnarray}
\Delta r=\Delta_1 \cos \omega t \cos N \theta+\Delta_2 \sin \omega t \cos (N+1) \theta,\\
\Delta \theta=\Delta_3 \cos \omega t \sin N \theta+\Delta_4 \sin \omega t \sin (N+1) \theta,
\end{eqnarray}
where the $\Delta_i$ are dimensionless numbers.

We assume that $\epsilon \ll 1$. As in the case of the Taylor sheet, we use no slip boundary conditions at the surface of the squirmer. The radial and  azimuthal components of the velocity field at the surface  of the squirmer are then given by:
\begin{eqnarray}
v_\rho(\rho=r,\theta=\phi)&=&\dot{r}=\epsilon R \omega (-\Delta_1 \sin \omega t \cos N \theta+\Delta_2 \cos \omega t \cos (N+1) \theta)],\label{bc3}\\
v_\theta(\rho=r,\theta=\phi)&=&r \dot{\phi}=R[1+\epsilon (\Delta_1 \cos \omega t \cos N \theta+\Delta_2 \sin \omega t \cos (N+1) \theta)]\nonumber \\
&& \times\epsilon \omega (-\Delta_3 \sin \omega t \sin N \theta+\Delta_4 \cos \omega t \sin (N+1) \theta).\label{bc4}
\end{eqnarray}
In polar coordinates the velocity components are related to the stream function $\psi$ by
\begin{equation}
v_\rho=\frac{1}{\rho}\dfrac{d\psi}{ d\theta},v_\theta=-\dfrac{d\psi}{ d\rho},
\end{equation}
We expand $\psi$, $Q_{\alpha \beta}$ and $U$  in $\epsilon$:
\begin{eqnarray}
\psi&=&\psi^{(1)} \epsilon+\psi^{(2)} \epsilon^2...\\
Q_{\alpha \beta}&=&Q^{(1)}_{\alpha \beta} \epsilon+Q^{(2)}_{\alpha \beta } \epsilon^2...\\
U&=& U^{(1)}\epsilon+ U^{(2)}\epsilon^2...
\end{eqnarray}
The boundary conditions \eqref{bc3} and \eqref{bc4} yield to first order in $\epsilon$
\begin{eqnarray}
\left. \dfrac{d\psi^{(1)}}{d\theta}\right| _{(R,\theta)}=R^2 \dfrac{d \Delta r}{dt}\label{bc5}\\
\left.\dfrac{d\psi^{(1)}}{d\rho}\right|_{(R,\theta)}=R \dfrac{d \Delta \theta}{dt}\label{bc6}
\end{eqnarray}
and to second order
\begin{eqnarray}
\left.\dfrac{d\psi^{(2)}}{d\theta}\right| _{(R,\theta)}=\left[ \dfrac{d\psi^{(1)}}{d\theta}-R \dfrac{d^2\psi^{(1)}}{d\rho d\theta}\right] _{(R,\theta)}\Delta r-
\left.\dfrac{d^2\psi^{(1)}}{ d\theta^2}\right| _{(R,\theta)}\Delta \theta\\
\left.\dfrac{d\psi^{(2)}}{d\rho}\right| _{(R,\theta)}=-R \Delta r\left[  \left.\dfrac{d^2\psi^{(1)}}{d\rho^2 }\right| _{(R,\theta)}+\dfrac{d\Delta \theta}{dt}\right] - \left. \dfrac{d^2\psi^{(1)}}{d\rho d\theta}\right| _{(R,\theta)}\Delta \theta.
\end{eqnarray}
The boundary conditions at infinity are given by
\begin{eqnarray}
\left.\bm{\nabla} \times (\psi^{(1)} \hat{\textbf{z}})\right|_{(\infty,\theta)}=U^{(1)} \cos \theta \hat{\bm{\rho}}-U^{(1)} \sin \theta \hat{\bm{\theta}}\\
\bm{\nabla} \times (\psi^{(2)} \hat{\textbf{z}})|_{(\infty,\theta)}=U^{(2)} \cos \theta \hat{\bm{\rho}}-U^{(2)} \sin \theta \hat{\bm{\theta}}
\end{eqnarray}
The first order terms in $\psi$ and $Q_{\alpha\beta}$ have the following form:
\begin{subequations}
	\begin{eqnarray}
	\psi^{(1)}&=&\sum_{s_1,s_2=\pm 1}\sum_{ n=N}^{N+1}\psi^{(1)}_0(\rho,n,s_1,s_2)\exp(s_1n \theta +s_2 \omega t)\\
	Q^{(1)}_{\rho\rho}&=&\sum_{s_1,s_2=\pm 1}\sum_{ n=N}^{N+1}Q^{(1)}_{\rho\rho0}(\rho,n,s_1,s_2)\exp(s_1n \theta +s_2 \omega t)\\
	Q^{(1)}_{\rho\theta}&=&\sum_{s_1,s_2=\pm 1}\sum_{ n=N}^{N+1}Q^{(1)}_{\rho\theta 0}(\rho,n,s_1,s_2)\exp(s_1n \theta +s_2 \omega t)
	\end{eqnarray}\label{firstord}
\end{subequations}

Solving Eq. \eqref{Qeqn1} to first order with the form of the first order solutions given in Eqs. \eqref{firstord}, we find
\begin{eqnarray}
Q^{(1)}_{\rho\rho0}&=&\dfrac{2  n s_1 \mu }{(i A- s_2 \omega \nu)}\left[ \dfrac{1}{\rho} \dfrac{d \psi^{(1)}_0}{d \rho}-\dfrac{1}{ \rho^2}\psi^{(1)}_0\right] \\
Q^{(1)}_{\rho\theta 0}&=&\dfrac{i\mu }{iA- s_2\omega\nu}\left[\dfrac{n^2 s_1^2}{\rho^2}\psi^{(1)}_0-\dfrac{1}{\rho}\dfrac{d \psi^{(1)}_0}{d \rho}+\dfrac{d^2 \psi^{(1)}_0}{d \rho^2} \right]
\end{eqnarray}
Taking the curl of Eq. \eqref{eqfb1} and then substituting the above values of $Q^{(1)}_{\rho\rho 0}$ and $Q^{(1)}_{\rho\theta 0}$ yields to first order
\begin{equation}
\left[ \rho^4 \dfrac{d^4}{d\rho^4}+2 \rho^3 \dfrac{d^3}{d\rho^3}-(1+2 n^2 s^2_1)\rho^2\dfrac{d^2}{d\rho^2}+(1+2 n^2 s^2_1)\rho\dfrac{d}{d\rho}+n^2 s_1^2(-4+n^2s^2_1)\right]  \psi^{(1)}_0=0
\end{equation}
The finite value solutions of the above equation have the form
\begin{equation}
\psi^{(1)}_0=A_1(n,s_1,s_2)\rho^{-n}+A_2(n,s_1,s_2)\rho^{2-n}
\end{equation}
where $A_i(n,s_1,s_2),i=1,2$ are the constants. With the boundary conditions \eqref{bc5} and \eqref{bc6}, we find
\begin{eqnarray}
\nonumber
\psi^{(1)}&=&\mathcal{A}_1(\rho) \sin \omega t \sin N \theta+\mathcal{A}_2(\rho) \cos \omega t \sin (N+1) \theta\\
\nonumber
Q^{(1)}_{\rho\rho}&=&\mathcal{A}_3(\rho)(\nu \omega \cos \omega t-A\sin \omega t)\cos N\theta+\mathcal{A}_4(\rho)(A \cos \omega t+\nu \omega\sin \omega t)\cos (N+1)\theta\\
\nonumber
Q^{(1)}_{\rho\theta}&=&\mathcal{A}_3(\rho)(\nu \omega \cos \omega t-A\sin \omega t)\sin N\theta+\mathcal{A}_4(\rho)(A \cos \omega t+\nu \omega\sin \omega t)\sin (N+1)\theta
\end{eqnarray}
where  
\begin{eqnarray}
\mathcal{A}_1(\rho)&=&\frac{ (\Delta_3-\Delta_1)N (\rho/R) ^2+\Delta_1(N-2)-\Delta_3N}{2 N}\left( \dfrac{\rho}{R}\right)  ^{-N}\\
\mathcal{A}_2(\rho)&=&\frac{(\Delta_2-\Delta_4)(N+1) (\rho/R) ^2+\Delta_4(N+1)-\Delta_2(N-1)}{2 (N+1)}\left( \dfrac{\rho}{R}\right)  ^{-N-1}\\
\mathcal{A}_3(\rho)&=&\frac{\mu   \left[-(N+1)(N-2)\Delta_1+N(N+1)\Delta_3+N(N-1)(\Delta_1-\Delta_3)(\rho/R) ^2\right]}{A^2+\nu ^2\omega^2}\left( \dfrac{\rho}{R}\right)  ^{-N-2}\\
\mathcal{A}_4(\rho)&=&\frac{\mu   \left[-(N-1)(N+2)\Delta_2+(N+1)(N+2)\Delta_4+N(N+1)(\Delta_2-\Delta_4)(\rho/R) ^2\right]}{A^2+\nu ^2\omega^2}\left( \dfrac{\rho}{R}\right)  ^{-N-3}
\end{eqnarray}
We use the above first order solutions in the second order forms of Eqs.  \eqref{eqfb1} and \eqref{Qeqn1} and solve these equations using Mathematica with the following result for the swimming speed to second order:
\begin{eqnarray}
U=U_0+\dfrac{1}{8}\epsilon^2 R\omega\left[\frac{2\nu\mu \mu_{\mathrm{eff}}\omega^2}{\eta_{\mathrm{eff}}(A^2+\nu^2\omega^2)} \right] \mathcal{G}
\end{eqnarray}
where
\begin{eqnarray}
\nonumber
\mathcal{G}&=&\frac{\Delta_1 \Delta_2 (N (N (N (2 N+11)+3)-26)-10)-\Delta_2 \Delta_3 (N (N (N (2 N+11)+10)-6)+3)}{(N+1) (N+2) (N+3)}\\&-& \frac{\Delta_4 (N+1) (\Delta_1 (N+1) (N+4) (2 N+1)+\Delta_3 (N (N+2) (2 N+5)-3))}{(N+1) (N+2) (N+3)}
\end{eqnarray}
and $U_0$ is the speed of the squirmer in the Newtonian fluid: 
\begin{equation}
U_0=\frac{1}{8} \epsilon^2 R \omega\left[ \Delta_4 (2 N+1) (\Delta_1+\Delta_3)-\Delta_2 (\Delta_3+\Delta_1 (2 N-3)-2 \Delta_3 N)\right] .
\end{equation}
One can see from the above expression for the swimming speed $U$ that the change in the swimming speed due to anisotropy is proportional to 
$2\nu\mu \mu_{\mathrm{eff}}\omega^2/\eta_{\mathrm{eff}}(A^2+\nu^2\omega^2)$
which is qualitatively similar to the result obtained for the Taylor sheet [see Eq. (19) of the main article].